\documentclass[12pt,epsf]{article}

 \usepackage{epsfig}
 \usepackage{times}
 
 \usepackage{amsmath}
 \usepackage{amssymb}

\newcommand{\be}{\begin{equation}}
\newcommand{\ee}{\end{equation}}
\def\({\left (}
\def\){\right )}
\def\[{\left [}
\def\[{\right ]}

\begin{document}
\begin{titlepage}
\bigskip
\rightline
\bigskip\bigskip\bigskip\bigskip
\centerline {\Large \bf {Gravitation and tunnelling: }}
\vspace{0.1cm}
\centerline {\Large \bf {Subtleties of the thin-wall approximation}}
\vspace{0.1cm}
\centerline {\Large \bf { and rapid decays}}
\bigskip\bigskip
\bigskip\bigskip

\centerline{\large Keith Copsey}
\bigskip\bigskip

\centerline{\em Perimeter Institute for Theoretical Physics, Waterloo, Ontario N2L 2Y5, Canada}

\vspace{0.3cm}

\centerline{\em kcopsey@perimeterinstitute.ca}
\bigskip\bigskip

\begin{abstract}
 I provide some simple physical arguments that, once gravitation and some subtleties are taken into account, rather broad classes of potentials result in instantons which tunnel relatively rapidly between perturbatively stable minima.   In particular, due to some previously unappreciated technical subtleties, the decay rates for instantons which may be well-described as thin-wall are much larger than the usual Coleman-de Luccia result.    I discuss with some level of rigor when the thin-wall approximation holds and clarify some misconceptions regarding the application of this approximation and its meaning.  I also point out potentials involving small differences between the maxima and minima generically decay relatively rapidly.  These two classes of potentials include those usually used to argue for the existence of a string landscape and in light of these results it is not clear that de Sitter vacua presently understood will generically be cosmologically long-lived.

\end{abstract}
\end{titlepage}

\baselineskip=16pt
\setcounter{equation}{0}

\section{Introduction: Some modest observations}

It has now become standard lore in string theory that there appears to exist a vast landscape of metastable vacua \cite{KKLT, BoussoPolchinski, Susskind:2003kw, Douglas:2006es}.  Despite the determined efforts of a number of capable individuals over more than a few years, there seems to be very little consensus as to how to compute probabilities in such a context and what, if anything, should be predicted about low energy physics.  Given its past history of subtle evasion of various generic problems, one might hope that string theory should avoid such a scenario and that the existence of a landscape of cosmologically long-lived vacua is an illusion.  While some authors have previously raised somewhat subtle issues that might interfere with the existence of a landscape, at least among de Sitter vacua, (see, e.g. \cite{DineLand}-\cite{BanksLand}) here I wish to point out once some confusions are unravelled and technical subtleties are understood, gravity can make tunnelling rates between perturbatively stable minima much faster than has been generally appreciated.  In the absence of gravity, one can easily estimate the decay rate of such vacua with a WKB approximation and with, for example, suitably high barriers make them long-lived.   As pioneered by Coleman and de Luccia \cite{CdL}, the effects of gravity on decay rates can be studied by finding (approximate) instantons describing the decay and calculating the rate
\be \label{dcyrate}
\Gamma = A e^{-B}
\ee
with
\be
B = S_E - S_b
\ee
where $S_E$ is the Euclidean action of the instanton, $S_b$ the Euclidean action of the original perturbatively stable ``background''\footnote{Note it is important to distinguish this state from the final state as the instanton is required to have asymptotics consistent with this original state but its is sometimes useful to consider the time reversed version of the instanton, as well tunnelling from true to false vacua (in the asymptotically de Sitter case).  The term ``background'' in this context does not imply any assumption one is perturbatively close to this state.}  state (e.g. false vacuum) while the prefactor $A$ is related to the functional determinant of the measure and, in the absence of special symmetries (e.g. fermionic zero modes), might be expected to be of order the volume of the instanton in Planck units.\footnote{To the best of my knowledge, the calculation of $A$ including gravitational backreaction remains an open problem.}  Provided $B$ is large, however, $A$ will only enter as a logarithmic correction and only be relevant (assuming it is not exactly zero) if one is calculating $B$ to extremely high precision \cite{DineNearlySusy}.

While gravity is frequently negligible for local considerations, the decay rate is related to the action and even very small local effects can cumulatively have a large effect on such global quantities.  One might, for example, suppose that gravational backreaction is unimportant as long as the peturbations to the potential are small compared to the Planck scale.  The most obvious counter-example to such an idea is the observation that one may formed a trapped surface, and hence a black hole, with matter or gravitational perturbations that are arbitrarily small in magnitude (e.g. out of radiation).  As I detail below, even in the limit of arbitrarily small potential differences it turns out gravitational  backreaction effects on bubble nucleation turn out to be of the same order of magnitude as the non-gravitational ones.  A more frequently made assertion is the idea that as long as the nucleated bubble is small compared to the scale of the cosmological background, as well as, presumably, large compared to the self-gravitational (i.e. Schwarszschild) radius, the effects of gravity on the action, and hence decay rate, are negligible.   One could make an entirely analogous argument in an asymptotically flat or asymptotically anti-de Sitter (AdS) context where it is easy to see the argument fails--the action is related to the energy (see e.g. \cite{HorowitzHawking}) which in turn is given by deviation of the metric from its asymptotic limit.  In other words, in an asymptotically AdS space the energy and action both reflect the presence of, for example, a planet even if the size of that planet is large compared to its Schwarzschild radius and small compared to the cosmological scale.

Note, however, one should not conclude from the above observations that gravitational backreaction is typically important in terrestrial laboratory experiments.   If gravitational backreaction is dominated by other matter contributions besides those involved in the tunnelling process then to a good approximation one may just calculate the decay rate in a fixed background metric given by these sources.   This is, of course, typical in tabletop experiments--the energy transitions involved between tunnelling between two metastable states is usually small compared to the rest energy of the remainder of the experiment, let alone the rest energy of the Earth.   Cosmologically, however, such localized matter distributions are typically negligible and most cosmological models involve at least some era dominated by a scalar field or other matter contributions which, given suitable effective potentials, may tunnel.

In fact, there are two classes of potentials that one can easily argue, once gravitational effects are properly included, will lead to rapid decays in the case the only significant matter fields may be modelled by a scalar and an effective potential.   To see this, start with the $d$-dimensional ($d \geq 3$) Euclidean action\footnote{I will discuss issues of surface terms and regulation of divergences below but the reader concerned about this point may simply regard the present discussion as applying to the compact instantons of the asymptotically de Sitter case where neither of these concerns arise.}
\be
S_E = -\kappa \int \sqrt{g} (R - \frac{1}{2} (\nabla \phi)^2 - V(\phi))
\ee
where $\kappa = (16 \pi G_d)^{-1}$ and I have chosen the overall sign, as usual, to ensure that the production rate of black holes is de Sitter space is not Planckian but rare \cite{Bousso:1996wy, Bousso:1996wz}.  Taking the trace of Einstein's equations one finds 
\be
R = \frac{1}{2} (\nabla \phi)^2 + \frac{d}{d-2} V(\phi)
\ee
and then the on-shell action may be written as
\be \label{act1}
S_E = -\frac{2 \kappa}{d-2} \int \sqrt{g} \, V(\phi)
\ee
Note the effect of gravity has had two rather remarkable effects--the on-shell action no longer has any gradient terms and higher values of the potential, presuming the backreaction effects on the metric do not dominate, make the action more negative not more positive.  To come to any definitive conclusion one must take backreaction into account properly--for pure de Sitter space, for example, with $V(\phi) = V_0$
\be \label{SdS}
S_E = -\frac{2 \kappa \Omega_{d-1} \sqrt{\pi}}{d-2} \Big((d-1) (d-2) \Big)^{d/2} \frac{\Gamma \Big( \frac{d}{2} \Big)}{\Gamma \Big(\frac{d+1}{2} \Big)} V_0^{1-d/2}
\ee
and the fact that a lower cosmological constant results in a larger instanton volume more than compensates for the explicit potential in (\ref{act1}).   To obtain more explicit results for the cosmological situation it is useful to restrict one's attention to the maximally symmetric instantons
\be
ds^2 = d\tau^2 + \rho^2(\tau) d\Omega_{d-1}
\ee
where $d\Omega$ is the metric on the unit $(d-1)$-sphere and $\phi(\tau)$ .  Then (\ref{act1}) becomes 
\be \label{act2}
S_E = -\frac{2 \kappa \Omega_{d-1}}{d-2} \int_{0}^{\tau_f} d\tau \rho^{d-1} V(\phi)
\ee
where $\tau$ is chosen so that $\rho(0) = 0$ in all cases and in the asymptotically dS case $\rho(\tau_f) = 0$ as well, while in the asymptotically flat or asymptotically AdS case $\tau_f \rightarrow \infty$ and $\rho$ diverges at large $\tau$.

One class of potentials where one should get relatively rapid decays might be termed  ``small-wall'', i.e. ones where the difference in potential from a relative maximum to the nearest minima is very small compared to the values of the potential (away from zero).  Simply via continuity, since the decay rate (\ref{dcyrate}) involves subtracting the action of the false vacuum background from this instanton, then one expects $\vert S_E - S_b \vert \ll \vert S_b \vert $ and for a sufficiently small potential difference the decay should become rapid.   More precisely, as the potential differences become small $\phi'$ never becomes large and so the effects of backreaction on the metric are small.  Then at leading order $\rho$ for the instanton is given by $\rho$ in the original vacuum and so $S_E \approx S_b$.   The fifth section will discuss these solutions in detail and validate these expectations.

First, however, I will discuss a class of instantons where  $\rho$ is roughly constant when $\phi$ is not near a relative minima--this is the ``thin-wall'' approximation made famous by Coleman and de Luccia \cite{CdL}.  In terms of the potential such instantons occur if the barrier is thin (in Planck units) and the height of the barrier is large compared to the difference between the minima or when the thickness of the barrier is of order one if the barrier height is large compared to the minima and the instanton starts sufficiently close, in a matter made precise below, to a minima.   Under these circumstances the action splits up to a region well described by the initial vacua, one well described by the final vacua, and a ``wall'' region inbetween where $\rho \approx \rho_0$ for some constant $\rho_0$.  For a potential of width $\Delta \phi$, the amount of (Euclidean) time the instanton spends in the wall region will be of order
\be \label{transittime}
\Delta \tau \sim \frac{\Delta \phi}{\sqrt{V_B}}
\ee
where $V_{B}$ is the potential barrier height.  (\ref{transittime}) will be justified in detail below but the result should be intuitive--if one is near the top one would expect a timescale to be set by the curvature at the top of the potential (${V''(\phi_{B})}^{-1/2}$) and in any case the dependence on the potential is fixed on dimensional grounds.   Consider the decay rate for transitions as $V_B$ becomes large for a fixed width $\Delta \phi$, or, as it usually phrased, as the tension of the wall becomes large.  The contribution of the wall to the action will be of order
\be
-\rho_0^{d-1} V_B \Delta \tau  \sim - \rho_0^{d-1} \Delta \phi \sqrt{V_B}
\ee
Since $\rho_0$ is continuous between the regions as $V_B$ becomes larger this term will eventually dominate the action and by making $V_B$ large enough the action can be made as negative as one likes.  In particular, this action can become more negative than the action for the background de Sitter space $S_b$ (\ref{SdS}) and the decay rate (\ref{dcyrate}) is exponentially enhanced instead of exponentially suppressed.

The only way to avoid such a conclusion is to assume $\rho_0$ becomes arbitrarily small (compared to the length scales associated with the true and false vacua) as $V_B$ becomes large.  However, physically as one increases the wall tension the bubbles become larger not smaller.   To be more precise, there are two possible classes of small $\rho_0$ bubbles.   In the first $\rho'$ is positive inside the bubble and one has a small volume of true vacuum, a wall where $\rho \approx \rho_0$, and a false vacuum region and $\rho_0$ is small compared to the length scale associated with the false vacuum.  Such bubbles are well described as ordinary flat-space bubbles and even in the asymptotically de Sitter setting one will locally have an approximately conserved energy\footnote{Globally, the definition of energy is rather ambiguous at best in an asymptotically de Sitter setting due to the lack of a global timelike killing vector.}.   Then, as usual, a decay must be described as an (approximate) zero energy solutions where the energy one gains by tunnelling to a lower energy vacuum precisely compensates for the energy costs of a wall of tension $T$ or
\be
\rho_0^{d-1} (V_F - V_T) \sim T \rho_0^{d-2}
\ee
where $V_F$ and $V_T$ are the potentials of the false and true vacuum, respectively, and hence
\be \label{smallbubblecrit}
\rho_0 \sim T/(V_F - V_T)
\ee
and hence large tension walls correspond to large $\rho_0$.  It should be emphasized that the above relationship breaks down when $\rho_0$ becomes comparable to the false vacuum scale in the asymptotically de Sitter case and in particular I will later point out in this case $\rho_0$ becomes comparable to the length scale of the false vacua in the limit of large tension.  

The second class of bubbles with small $\rho_0$ occurs if the true vacuum has a positive potential and one stays near the true vacuum until $\rho$ reaches a maximum and is decreasing as one encounters the wall.  In other words, in this asymptotically de Sitter setting where the instanton is topologically a sphere, almost the entire volume of the instanton is in the true vacuum.   The expert reader may well recognize such a scenario as impossible; as I will discuss later in detail in such a case the friction term in the field equation for $\phi$ becomes an anti-friction term in the wall and in the false vacuum region, $\phi$ never comes to rest and the instanton is necessarily singular.  It is simpler, however, to note that if such a solution did exist it would have a more negative action that the false vacuum background as its almost entire volume is in a lower cosmological constant (\ref{SdS}), as well as some wall region which only makes the action more negative, and again one would conclude $S_E - S_b < 0$ and the decay rate would be exponentially enhanced.  

 These observations contrasts with the usual Coleman-de Luccia \cite{CdL} result that (in four dimensions) in the limit of large tension the decay of a positive false vacuum to flat space occurs at a rate with
\be \label{CdLrate1}
B_{CdL} =  \frac{96 \pi^4}{\kappa^2 V_F}
\ee
where $V_F$ is the potential of the false vacuum and hence predicts such decay rates  are strongly suppressed  as $V_F$ becomes small (in Planck units).    As I will explain in detail below, this difference can be traced to a technical subtlety in the Coleman-de Luccia treatment \cite{CdL}, as well as the succeeding work by Parke \cite{ParkeBubbles}, where the method of evaluating the action implicitly assumes the amount of Euclidean time it takes to transverse the tunnelling solution is the same as for the false vacuum solution.  This is generically not true once one takes into account backreaction and using a slightly different evaluation process one can avoid this assumption.   One then recovers thin wall decays which are rapid not only for large potential barriers but, remarkably enough, generically for instantons well described as thin-wall.

It is worth emphasizing that while the ``wall'' in the thin-wall approximation can be arbitrarily thin, and will generically be so in the limit of very thin or very tall barriers, the thin-wall approximation does not mean one takes the wall to be infinitesimally thin or, equivalently, a brane with a true delta-function distribution of matter.  Such a delta-function solution actually corresponds to a naked curvature singularity; the presence of the energy density induces a step-function in $\rho'$ and consequently a divergence in, for example, the square of the Riemann tensor.   Such a configuration viewed as an exact, rather than merely an approximate, solution is also doubtful from the point of view of stability--any small perturbation breaking the symmetries ought to produce a black hole--as well as the usual quantum mechanical spreading effects.   

 If one takes any regular configuration with a bounded potential and shrinks the width of the barrier so the wall becomes arbitrarily thin, the tension of the wall becomes small, resulting in an arbitrarily small bubble.   If, on the other hand, one tried to make a true delta-function limit retaining a finite size wall and tension by making a potential barrier which is infinitely thin and high, despite the dubious physical admissibility of such a configuration, one can directly show, at least in the case of asymptotically de Sitter (dS) solutions, the resulting instanton is necessarily singular.  Intuitively this obstruction, explained in detail in subsection 3.6 below, arises from the fact as $\phi$ crosses the wall it picks up a non-zero velocity and once in the minimum region $\phi'$ can not go to zero in any finite period of time and, as a result, the instanton is singular.   In other words, at least in the asymptotically dS case, trying to take the true delta-function limit of a regular configuration either results in no bubble at all or a singular instanton.

If one takes the point of view, on the other hand, that such a delta-function treatment is only an approximation and the true configuration is actually regular if one cares to look at small enough scales, then the above argument applies and if the potential is large enough this wall can be the dominant contribution to the action. In particular, while $\rho$ can be found in both near vacuum regions with increasing accuracy as the barrier becomes thin by using the junction conditions, correctly calculating the on-shell action, and hence the tunnelling rate, requires one to include the contribution of the wall to the action, including gravitational effects.  Ignoring this contribution is equivalent to dropping a term in the action and one should not expect to obtain an accurate result.  In particular, if one ignored the wall contribution due to gravity it is not difficult, either graphically or analytically, to convince oneself, at least for asymptotically dS solutions, that the action for tunnelling solutions is necessarily larger than the background solution and only yields tunnelling rates that are exponentially suppressed.

After reviewing tunnelling including gravitation and collecting some useful general facts, I will discuss and justify in detail the conditions under which the thin-wall approximation holds and attempt to clarify some misconceptions regarding its application.  The fourth section presents the detailed evaluation of thin-wall decays for a variety of cases and the fifth section discusses the small-wall instantons.  With apologies to the expert reader (who may regard substantial parts of the next two sections as dwelling on facts which are well known), I have attempted to make this discussion as self-contained as possible in anticipation of the possibility that these results may be of interest to a rather broad audience, including members of both the cosmological and string communities.  It may be useful, however, to first summarize the key results:

\begin{itemize}

\item{Terms that are ignored in the traditional methods of evaluating the action for thin-wall instantons are much larger than one might naively expect and one obtains qualitatively different results once this omission is corrected; the usual Coleman-de Luccia approach fails to take into account the effect of backreaction on the total duration of the instanton and there is no limit of regular (asymptotically de Sitter) instantons extremizing the usual scalar action which results in a finite size bubble with a truly infinitesimally thin wall (hence allowing one to evaluate the action using only the junction conditions).}

\item{The decay rates for asymptotically de Sitter instantons (i.e. positive false vacuum potential) which are well-described by the thin-wall approximation, including the model potential of the KKLT construction \cite{KKLT}, are exponentially large, not exponentially small, with the possible exception of decays to sufficiently deep anti-de Sitter minima where multiple thin-wall instantons appear to exist.}

\item{The decay rate of asymptotically flat to asymptotically AdS minima via thin-wall instantons is also exponentially enhanced, presuming, as appears likely, one can ignore surface terms and the action does not suffer from any subtle cutoff ambiguities.}

\item{The decay of negative potential minima appears to violate a variety of theorems, as well as the AdS-CFT correspondence, and must (apparently) be absolutely forbidden, as has been well-appreciated previously in certain quarters.   It is still possible, however, to construct numerically instantons that naively appear to describe such a decay.}

\item{Potentials where the difference between a maximum and the minima is small compared to the values of the potential decay at a rate proportional to this potential difference.  Such instantons need not decay at exponentially enhanced rates or be well-described by the thin-wall approximation but if this potential difference is sufficiently small (e.g. a very dense Abbot washboard potential \cite{Abbotpot}) these instantons decay rapidly.}

\end{itemize}

\setcounter{equation}{0}

\section{Review of gravitational scalar instantons}

Consider the action
\be
S_E = -\kappa \int \sqrt{g} \Big( R - \frac{1}{2} (\nabla \phi)^2 - V(\phi) \Big)
\ee
with an S0(d) symmetric metric
\be
ds^2 = d\tau^2 + \rho^2(\tau) d\Omega_{d-1}
\ee
where $d\Omega_{d-1}$ is the metric on the unit $(d-1)$-sphere and $\phi(\tau)$.  With these assumptions the Einstein and field equations are equivalent to
\be \label{Ein1}
{\rho'}^2 = 1 + \frac{\rho^2}{(d - 1) (d - 2)} \Big(\frac{1}{2} {\phi'}^2 - V(\phi) \Big)
\ee
\be \label{Ein2}
\frac{\rho''}{\rho} = -\frac{{\phi'}^2}{2 (d - 1)} - \frac{V(\phi)}{(d - 1) (d - 2)}
\ee
and
\be \label{Feqn}
\phi'' + (d - 1) \frac{\rho'}{\rho} \phi' = V'(\phi)
\ee
As is well known, these equations are somewhat redundant--provided the constraint (\ref{Ein1}) is satisfied initially the evolution equations [(\ref{Ein2}), ( \ref{Feqn})] guarantee it will be at later times or if one prefers satisfying the Einstein equations [(\ref{Ein1}), ( \ref{Ein2})] will ensure the field equation (\ref{Feqn}) is satisfied, as can be easily checked by taking the time derivative of (\ref{Ein1}).   
Choosing $\tau$ so that $\rho(0) = 0$ (the instanton will be geodesically incomplete if it does not include $\rho = 0$), the instanton will be topologically a sphere for the asymptotically de Sitter case or a disk for asymptotically flat or anti-de Sitter cases--if asymptotically $V(\phi) > 0$, asymptotically $\rho'' < 0$ and $\rho$ goes through a second zero whereas $\rho \rightarrow \infty$ as $\tau \rightarrow \infty$ if asymptotically $V(\phi) \leq 0$.   For all regular solutions, (\ref{Ein1}) requires that as $\tau \rightarrow 0$
\be \label{rhobdycond}
\rho \rightarrow \tau
\ee
and (\ref{Feqn}) requires
\be \label{phibdy}
\phi \rightarrow \phi_0 + \mathcal{O}(\tau^2)
\ee
for some constant $\phi_0$.   In the case where the scalar field begins exactly in an extrema of $V(\phi)$ (i.e. $V'(\phi_0) = 0$) then the unique solution to (\ref{Ein1}-\ref{Feqn}) is $\phi(\tau) = \phi_0$ and
\be \label{rhosol}
\rho(\tau) = \frac{1}{\omega_0} \sin (\omega_0 \tau)
\ee
where
\be \label{omega03}
\omega_0^2 = \frac{V(\phi_0)}{(d-1) (d - 2)}
\ee
and in the case $\omega_0  = 0$ one takes the obvious limit (i.e. $\rho = \tau$).    Any instanton describing tunnelling starts and ends some distance away from the minima (or one would never leave this minima), although in, for example, the thin-wall approximation $\phi$ will begin quite near to a minima.   To characterize these near minima regions, combining the Einstein equations (\ref{Ein1}, \ref{Ein2})
\be \label{Eincom1}
V(\phi) = -\frac{(d-2)}{\rho^2} \Big[ (d-2) (\rho'^2-1) + \rho \rho'' \Big]
\ee
and so as long as $\phi$ remains in a neighborhood of $\tau = 0$ ($\rho(0) = 0$) and so $V(\phi) \approx V(\phi_0)$, for $\phi(0) = \phi_0$, the unique approximate solution to (\ref{Eincom1}) satisfying (\ref{rhobdycond}) is
\be
\rho \approx  \frac{1}{\omega_0} \sin (\omega_0 \tau)
\ee
Then as long as $V(\phi) \approx V(\phi_0)$ and $V'(\phi) \approx V'(\phi_0) + V''(\phi_0) (\phi - \phi_0)$ and the equation for $\phi$ (\ref{Feqn}) near $\phi_0$ becomes
	\be
	\phi'' + (d-1) \omega_0 \cot \omega_0 \tau \phi' \approx V'(\phi_0) + V''(\phi_0) (\phi - \phi_0)
	\ee
	which may be solved in terms of a hypergeometric function that generically diverges as $\tau \rightarrow \pi/\omega_0$
	\be \label{phinear1}
	\phi \approx \frac{V'(\phi_0)}{V''(\phi_0)} \Bigg[ F\Big(a_0, b_0, \frac{d}{2}, \sin^2 \frac{\omega_0 \tau}{2} \Big) - 1 \Bigg] + \phi_0
	\ee
	where
	\begin{eqnarray}
	a_0 &=& \frac{1}{2} \Big( d - 1 - \sqrt{(d - 1)^2 - V_3} \Big) \nonumber \\
	b_0 &=& \frac{1}{2} \Big( d - 1 +\sqrt{(d - 1)^2 - V_3} \Big) 
	\end{eqnarray}
	and
	\be
	V_3 \equiv 4 (d - 1) (d - 2) \frac{V''(\phi_0)}{V(\phi_0)}
	\ee
	although in the case $a_0 = -n_0$, for $n_0$ a natural number, or equivalently that
	\be
	V''(\phi_0) = -\frac{n_0 (n_0 + d -1)}{(d-1) (d-2)} V(\phi_0)
	\ee
	then (\ref{phinear1}) truncates to a finite polynomial.  Alternatively in terms of a convergent hypergeometric function, at least for $V_0 > 0$,
	\be \label{phinear2}
	\phi \approx \frac{V'(\phi_0)}{V''(\phi_0)} \Bigg[ \cos^{2-d} \frac{\omega_0 \tau}{2} \, F\Big(a_1, b_1, \frac{d}{2}, \sin^2 \frac{\omega_0 \tau}{2} \Big) - 1 \Bigg] + \phi_0
	\ee
	where
	\begin{eqnarray}
	a_1 &=& \frac{1}{2} \Big( 1 - \sqrt{(d-1)^2 - V_3} \Big) \nonumber \\
	b_1 &=& \frac{1}{2} \Big( 1 +\sqrt{(d-1)^2- V_3} \Big) 
	\end{eqnarray}
	It is also possible to express these answers in terms of Legendre functions, as has been noted before \cite{BanksJohnsonEtII}.  In numerical investigations of tunnelling instantons, it is fairly common for $\phi_0$ to be quite close to the minima, especially as one approaches the thin-wall approximation, and the above expressions become rather useful.

Even imposing the above boundary conditions at $\tau = 0$,  one must tune $\phi_0$ or the instanton will still not be regular at larger $\tau$.  In the asymptotically flat or anti-de Sitter  (AdS) cases, (\ref{Feqn}) admits $\phi$ solutions which diverge exponentially as $\tau \rightarrow \infty$ and the coefficients of these terms must be tuned to zero.  In the asymptotically de Sitter case (dS) one must choose $\phi_0$ so that $\rho$ hits its second zero just as $\phi'$ vanishes.  The simplest such dS solutions, and the ones I will focus on here, are those where the first nontrivial zero (i.e. away from $\tau = 0$) of $\phi'$ occurs when $\rho$ vanishes.  As long as the curvature of the potential at the relative maxima is sufficiently negative one is guaranteed such a solution exists for potentials with a positive false vacuum energy \cite{SteinhardtJensenI}. To see this note that if one starts $\phi_0$ too close to the false vacuum, the time scale for $\phi$ to leave this neighborhood and cross the barrier (as $\phi_0$ approaches the false vacuum the time it takes to leave the vincinity of the minima goes to infinity) exceeds the time scale for $\rho$ to reach its maxima and goes through a second zero while $\phi'$ is still positive.   Such a solution is sometimes referred to as an overshoot. 

On the other hand, if $\phi_0$ starts very near the maximum of the potential at $\phi = \phi_B$ then as long as $\phi$ is near this maximum
\be \label{rhotop}
\rho \approx \frac{1}{\omega_B} \sin \omega_B \tau
\ee
where 
\be
\omega_B^2 = \frac{V (\phi_B)}{(d-1) (d - 2)}
\ee
and then the field equation in a neighborhood of this maximum is approximately
\be \label{phitopeqn}
\phi'' + (d-1) \omega_B \cot(\omega_B \tau) \phi' \approx V''(\phi_B) (\phi- \phi_B)
\ee
(\ref{phitopeqn})  admits the simple solution satisfying the desired regularity conditions 
\be \label{simplesol}
\phi = -\frac{C_0}{\omega_B} \cos \omega_B \tau + \phi_B
\ee
for some constant $C_0$ provided that
\be
V''(\phi_B) = - d \omega_B^2
\ee
Alternatively (\ref{simplesol}) can be derived as the first truncated polynomial from the hypergeometric general solution (\ref{phinear1}).
If
\be \label{Vbd}
V''(\phi_B) < -d \omega_B^2 = -\frac{d \, V(\phi_B)}{(d - 1) (d - 2)}
\ee
then $\phi'$ will go through a second zero before $\rho$ reaches its second zero at time $\tau = \pi/\omega_B$--such a solution is referred to as an undershoot.  Then provided the curvature near the maxima is sufficiently negative, one has solutions where $\rho$ reaches a zero before $\phi'$ and vice versa.  Via continuity there must be a $\phi_0$ where $\phi'$ and $\rho$ reach zero simultaneously.  Under rather mild additional assumptions, one can also argue such a solution is unique  and if $V''(\phi_B) > -d \omega_B^2$ there are no regular instantons  of this type \cite{SteinhardtJensenI}.   

For generic potentials with a barrier of width  $\Delta \phi$ 
\be
V''(\phi_B) \lesssim -\frac{V(\phi_B)}{\Delta \phi^2}
\ee
and hence (\ref{Vbd}) will typically provide no obstacle for thin barriers.  On the other hand, generic potentials with wide barriers ($\Delta \phi \gg 1$) will violate (\ref{Vbd}) and one only has to be concerned with a Hawking Moss \cite{HawkingMoss} instability.  Due to this observation, as well as to avoid some technical complications which will become apparent later, I will limit my attention here to $\Delta \phi \lesssim \mathcal{O}(1)$.     

If one defines a Euclidean energy for the scalar field as
\be \label{Edef}
E_0 = V(\phi) - \frac{{\phi'}^2}{2}
\ee
then if the ``friction'' term in (\ref{Feqn}) is relatively small
\be \label{Vprimebig}
\Big \vert  \frac{\rho'}{\rho} \phi' \Big \vert \ll \vert V'(\phi) \vert
\ee
then
\be
\phi'' \approx V'(\phi)
\ee
and $E_0$ will be approximately conserved.   This will occur, for example, away from the extrema of the potential if $\rho$ is relatively large.   Note as long as $\rho'$ is positive this friction term damps the acceleration of $\phi$ but if $\rho'$ becomes negative, as it always eventually does in the asymptotically de Sitter case, it will act as an anti-friction term and accelerates $\phi$.

It is also useful to rephrase the above statements regarding energy in a slightly different language.  Multiplying (\ref{Feqn}) by $\phi'$ and treating the resulting equation as a first order ordinary differential equation for ${\phi'}^2$ and demanding the instanton is regular at $\tau = 0$ one finds
\be \label{phi2}
\frac{{\phi'}^2}{2} = V(\phi) - \frac{2 (d - 1)}{\rho^{2 (d - 1)}(\tau)} \int_0^{\tau} ds \, V(\phi(s)) \, \rho^{2d - 3}(s) \rho'(s)
\ee 
In the asymptotically de Sitter case note that as $\rho$ will go to another zero at some finite time $\tau_f$, $\phi'(\tau_f)$ will diverge unless
\be \label{consistcondit}
\int_0^{\tau_f} ds V(\phi) \rho^{2 d - 3} \rho' = 0
\ee
Given (\ref{phi2}) it is straightforward to show that if $V_0 \leq V(\phi) \leq V_1$ that if $\rho'$ is non-negative
\be \label{philim}
V(\phi) - V_1 \leq \frac{{\phi'}^2}{2} \leq V(\phi) - V_0
\ee
Likewise, using (\ref{consistcondit}), if $\rho'$ is negative one again finds the same condition (\ref{philim}).   One might object that the latter comment assumes one has dS asymptotics, but if $\rho'$ becomes negative it stays negative after that point; a transition from negative $\rho'$ to positive $\rho'$ is only conceivably possible, via (\ref{Ein2}), if $V(\phi) < 0$ but $\rho'^2 \geq 1$ in any region where $V(\phi) < 0$ via the constraint (\ref{Ein1}) and so such a transitiion is impossible.  Hence negative $\rho'$ occurs only in asymptotically dS solutions.   While the left hand side of (\ref{philim}) is trivial, the right hand side indicates that ${\phi'}^2$ is bounded above by the potential difference--in other words, for regular solutions anti-friction, if present, can only compensate for the effects of friction.  In terms of energy (\ref{philim}) implies
\be \label{Ebds}
V_0 \leq E_0 \leq V(\phi)
\ee

Via the constraint (\ref{Ein1}), the bounds on $\phi'^2$ (\ref{philim}) also implies a bound on $\rho'^2$
\be \label{rhobound}
1 - \frac{\rho^2 V(\phi)}{(d-1) (d - 2)}  \leq \rho'^2 \leq 1-\frac{\rho^2 V_0}{(d-1) (d - 2)} 
\ee
Then as long as the potential is everywhere non-negative, $V_0 \geq 0$ and hence  $\rho'^2 \leq 1$. On the other hand, in any region of negative potential, via the constraint (\ref{Ein1}), $\rho'^2 > 1$ (except, of course, at the endpoint when $\rho = 0$ and $\rho'^2 = 1$).  However, for asymptotically dS solutions, as long as the true vacuum does not become too negative, $\rho'^2$ is at most of order one.  To see this note that the extrema of $\rho'^2$, aside from the minima at $\rho' = 0$ and the boundary conditions $\rho'^2 = 1$ when $\rho = 0$, occurs, if at all, when $\rho'' = 0$ or, via (\ref{Ein2}), at $t = t_c$ when
\be
V(\phi(t_c)) = -\frac{d-2}{2} \phi'^2(t_c)
\ee
Then if there is a region of the instanton where $V(\phi) < 0$
\be
\rho'^2 \leq 1 - \frac{\rho^2 V(\phi(t_c))}{(d-2)^2} \leq 1 - \frac{\rho^2 V_T}{(d-2)^2}
\ee
where $V_T$ is the potential of the true vacuum.   Hence ${\rho'}^2$ can potentially only become large if $\rho$ is large enough that $\rho^2 \vert V_T \vert \gg 1$. On the other hand, as long as the potential at the false vacuum $V_F$ is positive, then
\be \label{rhomaxbd}
\rho \lesssim V_F^{-1/2}
\ee
To see this, consider the instanton starting on the false vacuum side of the potential at $V(\phi) = V_0$.  If $\phi$ stayed nearly at rest near the false vacuum indefinitely, $\rho$ would reach a maximum scale set by $V_F$ (\ref{rhosol}).  On the other hand, (\ref{Ein2}) shows a nonzero ${\phi'}^2$ and/or an increasing potential make $\rho''$ more negative and hence reduce the maximum value of $\rho$ (recall the boundary condtions force $\rho'(0) = 1$).   One might worry that $\phi$ could travel to a value of the potential lower than $V_0$, and even $V_F$, before the instanton ends and invalidate the above argument.  However, at such a point (\ref{philim}) becomes impossible and hence $\rho'$ is necessarily negative and $\rho$ has already passed through its maximum.  Hence (\ref{rhomaxbd}) has been justified.  Then, unless $V_T \ll -V_F$, 
\be \label{rhogenbd}
\rho'^2 \lesssim 1
\ee
In fact, (\ref{rhogenbd}) appears to be true rather generically even if $V_T \ll -V_F$; it is straightforward to numerically find instantons with $V_T < 0$ (and where the instanton begins rather close to $V_T$) where $\vert V_T \vert$ is large  compared to $V_F$ and even the barrier maxima $V_B$ and in all the examples I have examined (\ref{rhogenbd}) holds.  On the other hand, there does not appear to be an obvious analytic argument proving (\ref{rhogenbd}) generically (and in the absence of such a bound it appears difficult to prove much for generic potentials regarding the validity of the thin-wall approximation) so the arguments of the next section regarding the validity of the thin-wall approximation will be strictly justified if  $V_F > 0$ and $V_T \gtrsim -V_F$, although many of the observations, as well as possibly the conclusions,  may apply more broadly.

If one specializes to thin barriers, however, one need not make the above assumptions to make progress.  If (\ref{rhogenbd}) fails and $\rho'^2 \gg 1$ then using the constraint  (\ref{Ein1}) it is not difficult to show that unless $V(\phi) \ll -\phi'^2$
\be \label{rhophibd}
\rho \sim \rho_0 E^{C_0 (\phi - \phi_0)}
\ee
where $\rho_0$ and $\phi_0$ are constants  (with the obvious interpretation) and $C_0$ is a constant of order one.  To be more precise, there are upper and lower bounds on $\rho$ in any region $\rho'^2 \gg 1$ where each bound has a constant $C_0$ which is of order one.  If, on the other hand, $V(\phi) \ll -\phi'^2$, it is easy to show from the Einstein equations (\ref{Ein1}, \ref{Ein2}) that $V(\phi)$ is approximately constant and one is not crossing the barrier at all.  Then as long as the barrier is thin $\phi - \phi_0 \ll 1$and $\rho$ is approximately constant in any region where $\rho'^2 \gg 1$.   On the other hand, if $\rho'^2 \gg 1$ in some region of width order one (in terms of $\phi$) $\rho$ will generically suffer an order one fractional change.

Given an instanton, one can then evaluate the on-shell (bulk) action (\ref{act2}) 
\be \label{act4}
S_E = -\frac{2 \kappa \Omega_{d-1}}{d-2} \int_{0}^{\tau_f} d\tau \rho^{d-1} V(\phi)
\ee
and calculate the tunnelling rate (\ref{dcyrate})
\be 
\Gamma = A e^{-B}
\ee
where $B$ is given by the difference between the instanton action and the background solution
($B = S_E - S_b$).   In the asymptotically de Sitter case the instanton is topologically spherical and hence (\ref{act4}) is finite and there are no boundary terms to be considered.  On the other hand, for asymptotically flat or AdS solutions as a matter of principle surface terms should be added to the action to ensure a well-defined variational principle\footnote{To produce the equations of motion for a given action one must perform integrations by parts and doing so results in a series of surface terms.  One must add surface terms to the bulk action such that the variation of these surface terms cancels the terms produced by integrating by parts or the equations of motion will not extremize the action-- in other words, one does not have a well-defined variation principle.}  and to make the action finite in the AdS case (although I do not know that this procedure has been carried out in the case of  cutoffs that appear to be natural cosmologically) \cite{KrausVijay}-\cite{MannMarolfVirmani}.  On the other hand, the bulk action appears to be perfectly finite in the asymptotically flat case--it is easy to check using [(\ref{Ein1})-(\ref{Feqn})] that asymptotically (normalizable) corrections to $\phi$ (as well as $\rho$) are exponentially suppressed as functions of $\tau$ and so the bulk action (\ref{act4}) is convergent.   Generically it does not appear obvious that surface terms could not contribute finite terms to the tunnelling rate and if this does occur that the result is independent of the cutoff procedure used (or cosmologically which is ``preferred'' if the terms are not cutoff independent).  Likewise if the above expectations fail and the action including surface terms is not finite, and hence one must introduce some additional regularization procedure, it is not obvious all such regularization procedures yield the same finite $B$.

In the absence of a detailed investigation of the above, it seems difficult to make rigorous predictions regarding tunnelling for the asymptotically flat or asymptotically AdS cases.   As a result of these considerations, as well as some technical difficulties mentioned later, I will largely focus on tunnelling from asymptotically de Sitter spaces.  I should note though that in the absence of evidence to the contrary it appears reasonable to expect the above considerations will be unimportant for the asymptotically flat case but rule out asymptotically AdS decays.  In the asymptotically flat case, as noted above, the (normalizable) perturbations of $\phi$, and hence presumably the metric, are exponentially  suppressed as functions of $\tau$ and hence $\rho$.   Surface terms, on the other hand, typically pick out out subleading contributions to the metric and matter fields that are only power-law suppressed as functions of the volume under consideration.  Hence, it would appear likely the resulting surface terms vanish in the asymptotically flat case.  It is also worth noting the usual positive energy theorems \cite{WSY} do not automatically forbid decays from flat space into a negative potential vacuum; once $V(\phi)$ goes negative the weak energy condition is generically violated and unless one is forced to go over a sufficiently high barrier before getting to a negative energy region one expects the spacetime to be unstable.  One should also not be misled into thinking supersymmetry guarantees the stability of such solutions; if the spacetime were exactly flat it would be supersymmetric and the decay rate would vanish (thanks to a fermionic zero mode) but the decays described always occur when $\phi$ fluctuates some distance away from the exact minima (typically into some region with positive potential energy where supersymmetry is broken).

The decay of an asymptotically AdS space for the potentials typically of interest, on the other hand, would seem to contradict not only the AdS-CFT correspondence \cite{BanksTasi} but a series of positive energy proofs for asymptotically AdS spacetimes showing the unique zero energy state is empty AdS \cite{PosEnergyAdS}.  To be more precise, these results apply provided the curvature of the potential asymptotically is not too negative compared to the AdS scale--a criterion ($m^2 \geq (d-1) \Lambda/{4 (d-2)}$) known as the Breitlohner-Freedman bound \cite{BF}.  Any potentials violating this bound generally yield highly unstable solutions although, of course, in the tunnelling context one is typically interested in potentials which monotonically increase from the minima to some barrier and $V''(\phi_{min}) > 0$ (and hence correspond to stable spacetimes).  It is not hard to numerically find an instanton satisfying (\ref{Ein1}-\ref{Feqn}) which interpolates between two AdS minima with a potential (respecting the Breitlohner-Freedman bound) so apparently considerations of the above kind (or some other complications I have not thought of) forbid the interpretation of such instantons as decays.

On the other hand, asymptotically de Sitter solutions can not only decay to a lower potential minima but, as has been noted previously \cite{LeeWeinberg}, tunnel to a higher potential minima.  However, this rate is necessary small compared to the decay rate.  Denoting the actions for the false vacuum, true vacuum, and instanton interpolating between them as $S_F$, $S_T$, and $S_I$ respectively, note that \cite{BanksJohnsonI} the same instanton can be used to calculate either the rate
\be
\Gamma_{F \rightarrow T} = A e^{-(S_I - S_F)}
\ee
or
\be
\Gamma_{T \rightarrow F} = A e^{-(S_I - S_T)}
\ee
and hence
\be
\Gamma_{T \rightarrow F} = e^{S_T - S_F}\Gamma_{F \rightarrow T} 
\ee
and hence, since the action for a de Sitter spacetime with $V(\phi) = V_0$ is given by
\be 
S_0 = -\frac{2 \kappa \Omega_{d-1} \sqrt{\pi}}{d-2} \Big((d-1) (d-2) \Big)^{d/2} \frac{\Gamma \Big( \frac{d}{2} \Big)}{\Gamma \Big(\frac{d+1}{2} \Big)} V_0^{1-d/2}
\ee
and becomes more negative with decreasing $V_0$,
\be
\Gamma_{T \rightarrow F}  \ll \Gamma_{F \rightarrow T} 
\ee
The above argument does not, however, immediately generalize to transitions between spaces with different kinds of asymptotics (e.g. flat to dS transitions) as the instanton is conventionally required to maintain the asymptotics of the original vacuum.

\setcounter{equation}{0}

\section{The application and limitations of the thin-wall approximation}

\subsection{Summary}

As previously mentioned, the thin wall approximation applies when $\rho$ is approximately constant when $\phi$ is not near a relative minima.  This will be achieved when the instanton starts rather close to a minima of the potential and exits this region only when $\rho$ is sufficiently large that the amount $\rho$ changes when $\phi$ is not near a minima is small compared to this ``initial'' value.  Such large $\rho$ also ends up implying $\phi$ has an approximately conserved energy $E_0$ (\ref{Edef}) away from the endpoints of the instanton and extrema of the potential.  After reviewing the conditions under which one has a conserved $E_0$, I show the time $\phi$ takes to cross a barrier of potential maximum $V_B$ and width $\Delta \phi$ is of order
\be
\Delta \tau \sim  \frac{\Delta \phi}{\sqrt{V_B}}
\ee
presuming the width of the barrier $\Delta \phi$ is not large in Planck units ($\Delta \phi \lesssim \mathcal{O}(1)$)  and that if the true vacuum potential $V_T$ is negative its magnitude is at most of order the barrier height.   Provided also that  ${\rho'}^2$ is at most of order one (\ref{rhomaxbd}), as will be assured if $V_T \gtrsim - V_F$ where $V_F$ is the false vacuum potential,  then the change in $\rho$ as $\phi$ travels over the barrier is at most of order $\Delta \tau$.  Then under these assumptions if $\rho \gg \Delta \tau$ at the point where $\phi$ leaves the neighborhood of a minima, the thin-wall approximation will be reliable.  

For thin barriers ($\Delta \phi \ll 1$) the solution will be well described as thin-wall if
\be \label{thinwallcond1}
V_B \gg V_F - V_T
\ee
In the case that $\vert V_T \vert \ll V_B$ if $V_T < 0$ then (\ref{thinwallcond1}) is not just a sufficient but a necessary condition for a thin-wall instanton.  For $\Delta \phi = \mathcal{O}(1)$ the thin-wall approximation will hold only if $V_B \gg V_F$ and further $\phi$ at one endpoint of the instanton begins sufficiently close, in a sense made precise below, to a minima.  I review a numerical procedure to check the later criterion for any potential of interest, in particular verifying the KKLT model potential \cite{KKLT} is well-described by the thin-wall approximation.  I also illustrate the importance of this criterion for non-thin barriers by showing a series of instantons not well-described as thin-wall in potentials where $V_B \gg V_F \approx V_T$.  

I finish the section by attempting to clarify some misconceptions regarding the thin-wall approximation, including the assertion that it implies the friction term in the field equation (\ref{Feqn}) may be simply dropped and the statement that the intermediate ``wall'' may be treated as infinitesimally thin compared to any other scale in the problem.  The justification of the above assertions appears to be necessarily somewhat technical and the reader eager to get to the results for thin and small-wall decays and content to accept the above statements, or the expert already familiar with them, should be able to skip to the next section.

\subsection{$\rho$ and conservation of energy}

Given the above results $\phi'^2$ is always bounded (\ref{philim}) and provided $\rho'$ is also bounded (which, as discussed above, follows if $V_T \gtrsim - V_F$) then if $\rho$ is sufficiently large at least away from the extrema of the potential the friction term in the field equation  (\ref{Feqn}) becomes small
\be
\Big \vert \frac{\rho'}{\rho} \phi' \Big \vert \ll \vert V'(\phi) \vert
\ee 
and so one has an approximately conserved energy
\be
E_0 = V(\phi) - \frac{{\phi'}^2}{2}
\ee
It is useful to define the difference between the potential at the maximum $V_B$ and the potential at the false vacuum $V_F$
\be
\Delta V_0 = V_B - V_F
\ee
and the difference between $V_B$ and the value of the potential at the true vacuum $V_T$ as
\be
\Delta V_1 = V_B - V_T
\ee
the width between the true vacuum and the barrier
\be
\Delta \phi_0 = \vert \phi_1 - \phi_0 \vert
\ee
where $V(\phi_0) = V_T$ and $V(\phi_1) = V_B$ and 
the width between the barrier and the false vacuum
\be
\Delta \phi_1 = \vert \phi_2 - \phi_1 \vert
\ee
where $V(\phi_2) = V_F$.  Note that
\be
\frac{\Delta V_1}{\Delta V_0} = \frac{V_B - V_F}{V_B-V_T} = \frac{1 - \frac{V_F}{V_B}}{1-\frac{V_T}{V_B}}
\ee
and hence unless $V_F \approx V_B$ (in which case $V(\phi) \approx V_B$ until $\phi$ crosses the maxima of the barrier to a good approximation) or $V_T \ll -V_B$ (in violation of the above assumption) then 
\be \label{V1s}
\Delta V_1 \sim \Delta V_0 \sim V_B
\ee

Away from the extrema of the potential, for an order of magnitude estimate one may approximate $V'(\phi)$ by its average
\be
V'(\phi) \sim \frac{1}{\Delta \phi_i} \int d \phi \, V'(\phi) = \pm \frac{\Delta V_i}{\Delta \phi_i}
\ee
with the sign and $i$ ($i = 0, 1$) chosen depending on what side of the relative maximum $\phi$ is on.  Then, given the bound on $\phi'^2$ (\ref{philim}) and presuming $\rho'^2 \lesssim 1$ (\ref{rhogenbd})
\be
\Big \vert \frac{\rho'}{\rho} \phi' \Big \vert \lesssim \frac{ \sqrt{ V(\phi) - V_0}}{\rho} \lesssim \frac{\sqrt{\Delta V_i}}{\rho}
\ee
using (\ref{V1s}) and excluding the case $V_F \approx V_B$ since in that case it makes little sense to talk about a region on the false side of the vacuum not near the extrema.  Then if
\be \label{rhofrict1}
\rho \gg \frac{\Delta \phi_i}{\sqrt{\Delta V_i}}
\ee
then 
\be
\Big \vert \frac{\rho'}{\rho} \phi' \Big \vert \ll \frac{\Delta V_i}{\Delta \phi_i} \sim \vert V'(\phi) \vert
\ee
and one will have an approximately conserved energy.

As long as $V(\phi)$ is near its minimum value, and in particular if $\phi$ is near the true vacuum, then via the bound on the energy (\ref{Ebds})
\be
V_0 \leq E_0 \leq V(\phi)
\ee
then $E_0 \approx V_0$, regardless of the size of $\rho$.  On the other hand, the energy can be substantially different from the potential as $\phi$ begins to approach the false vacuum.  While very close to its endpoint on the false side of the vacuum $\phi'^2$ ultimately goes to zero, either via regularity in the de Sitter case or as part of the boundary conditions in the asymptotically flat or AdS cases, $\phi'^2$ can become relatively large near this endpoint and change the energy substantially.  In particular, using (\ref{phinear2}) if $\phi$ spends enough time near the false vacuum that $\rho$ reaches a maximum and decreases until $\rho \ll \omega_0^{-1}$ (where $\omega_0$ is, as before, (\ref{omega03}) with $V(\phi_0)$ approximately the false vacuum potential)
\be
\phi' \approx \frac{(d-2) (\phi - \phi_0)}{\rho}
\ee
and, for sufficiently small $\rho$, $\phi'^2$ can be comparable to the false vacuum potential even while $\phi - \phi_0$ remains small.  As discussed below, provided the barrier is either sufficiently high or narrow such $\rho$ can still be large enough that the thin-wall approximation will hold.  
Likewise near the potential maximum, since $V'(\phi)$ goes to zero, the friction term will eventually dominate in some neighborhood of the maximum no matter how large $\rho$ is.   As discussed in the next subsection, depending on the barrier height and width, as well as the slope of $\rho$ near the maximum, the change in $E_0$ can either be significant or negligible.   The fact that there are regions where $E_0$ is generically non-conserved even if $\rho$ is large away from the extrema should be no surprise; at the endpoints of the instanton $\phi'^2 \rightarrow 0$ and in the event that these endpoints are at unequal values of the potential, as they generically will be if these endpoints are near the potential minima, then $E_0$ must interpolate between these different values.

\subsection{Barrier crossing time}

Away from any extrema unless the friction term in (\ref{Feqn}) is dominant then, as an order of magnitude estimate
\be
\phi'' \sim V'(\phi) \sim \frac{\Delta V_i}{\Delta \phi_i}
\ee
Note in terms of finding the time any regular instanton spends away from the minima one might as well consider $V'(\phi)$ to be positive (simply reverse the direction of time and increasing $\phi$ if it is not), as above.  Then
\be \label{intphi1}
\phi \sim \frac{\Delta V_i}{2 \Delta \phi_i}(\tau - \tau_0)^2 + \phi'(\tau_0) (\tau - \tau_0) + \phi (\tau_0)
\ee
where $\tau_0$ is the ``initial'' time near the bottom of the barrier when $V'(\phi(\tau_0))  \sim \Delta V_i/{\Delta \phi_i}$. Recalling $\phi'(\tau_0) \leq \sqrt{2 (V(\phi(\tau_0)) - V_0)} \lesssim \sqrt{\Delta V_i}$, one quickly finds by time $\tau_1$
\be \label{tlapse1}
\tau_1 - \tau_0  \sim \frac{\Delta \phi_i}{\sqrt{\Delta V_i}}
\ee
$\phi$ has traveled a distance of order $\Delta \phi_i$ and acquires a velocity near the maximum of order $\sqrt{\Delta V_i}$.  

If it happens to be true that $\rho$ in this regime is large enough so that $E_0$ (\ref{Edef}) is approximately conserved, as discussed above, one can obtain a more precise estimate by noting
\be \label{Etime1}
\phi' = \sqrt{2} \sqrt{V(\phi) - E_0}
\ee
and hence in a time
\be \label{Etime2}
\tau_1 - \tau_0 = \int_{\phi_0}^{\phi_1} \frac{d \phi}{\sqrt{2} \sqrt{ V(\phi) - E_0}}
\ee
one will have traveled from near the false vacuum to near the top of the barrier with an anologous expression for the other side of the barrier.  (\ref{Etime2}) is, however, not entirely transparent and one can also estimate this quantity by multiplying both sides of (\ref{Etime1}) by $V'(\phi)$ with the result that
\begin{eqnarray} \label{Etime3}
\sqrt{2} \sqrt{V(\phi_1) - E_0} &-& \sqrt{2}  \sqrt{V(\phi_0) - E_0} = \int_{\phi_0}^{\phi_1} d\phi \frac{V'(\phi)}{\sqrt{2} \sqrt{V(\phi) - E_0}} \nonumber \\
&=& \int_{\tau_0}^{\tau_1} dt V'(\phi) \sim \frac{\Delta V_i}{\Delta \phi_i} (\tau_1 -\tau_0)
\end{eqnarray}
and since the left hand side of (\ref{Etime3}) is of order $\sqrt{V_B}$
\be
\tau_1 - \tau_0 \sim  \frac{\Delta \phi_i}{\sqrt{\Delta V_i}}
\ee
as before.

On the other hand, let us suppose that the friction term is dominant in some region away from the extrema of the potential or, equivalently that
\be
\Big \vert \frac{\rho'}{\rho} \phi' \Big \vert \gg V'(\phi) \sim \frac{\Delta V_i}{\Delta \phi_i}
\ee
As before, given $\rho'^2 \lesssim \mathcal{O}(1)$ and $\phi'^2 \leq 2 \Delta V_0$ (\ref{philim}), then friction domination requires (\ref{Ein1})
\be
\rho^2 \ll \frac{\Delta V_0}{(\Delta V_i)^2} (\Delta \phi_i)^2
\ee
and, again excluding the case $V_F \approx V_B$, hence
\be
\rho \ll \frac{\Delta \phi_i}{\sqrt{\Delta V_0}}
\ee
Then subsequently
\be
\rho^2 \phi'^2 \ll (\Delta \phi_i)^2 \lesssim \mathcal{O}(1)
\ee
and
\be
\rho^2 \vert V(\phi) \vert \ll \frac{\vert V(\phi) \vert}{\Delta V_0} (\Delta \phi_i)^2  \lesssim \mathcal{O}(1)
\ee
 since, as one can prove with a few lines of algebra, $\vert V(\phi) \vert \lesssim \Delta V_0$ (unless $V_T \approx V_B$, in which case one will be in the later described small-wall approximation) and subsequently in any region of friction domination
 \be
 \rho'^2 \approx 1
 \ee
 Then at most the amount of time spent in a friction dominated region is
 \be
 \Delta \tau_1 \lesssim \rho  \ll \frac{\Delta \phi_i}{\sqrt{\Delta V_0}}
 \ee
and consequently the change in $\phi$ in such a region is
\be
\Delta \phi = \int d\tau \phi' \lesssim \Delta \tau_1 \sqrt{\Delta V_0} \ll \Delta \phi_i
\ee
and hence, given the above assumptions, if there is any region of friction domination (away from the extrema of the potential) it occurs only in a very small region compared to the width of the barrier and lasts only a relatively short period of time.  Conversely, away from the extrema of the potential as long as $\rho$ is sufficiently large the instanton will not be friction dominated.

Now turning to the amount of time $\phi$ spends near the maximum of the potential, it is useful to rewrite the field equation (\ref{Feqn}) as
\be \label{phimax}
\frac{\phi'^2}{2} = V(\phi) - V(\phi(\tau_0)) + \phi'(\tau_0)^2- (d - 1) \int_{\tau_0}^{\tau} ds \frac{\rho'}{\rho} \phi'^2
\ee
Since, as previously noted, as $\phi$ approaches the maxima of the potential  $\phi' \sim \sqrt{\Delta V_i} \sim \sqrt{V_B}$ and given $\rho'^2 \lesssim 1$, then even in the most dissipative case it takes a time interval of order $\rho$ to result in a significant reduction of $\phi'$.  In other words, near the maximum of the potential,  $\phi'^2 \sim V_B$  at least until $\tau - \tau_0 \sim \rho$. Then in any time interval $\Delta \tau_2 \ll \rho$, $\phi$ will cover a distance
\be
\delta \phi = \int d\tau \phi' \sim \sqrt{V_B} \Delta \tau_2
\ee
and in a time
\be
\Delta \tau_2 \sim \frac{\delta \phi}{\sqrt{V_B}}
\ee
where the width of the maximum region is $\delta \phi$, $\phi$ would have crossed this region.   Hence if $\rho \gg \delta \phi/{\sqrt{V_B}}$ near the maximum of the potential, the barrier will
be crossed in a time of order $\delta \phi/{\sqrt{V_B}}$.  On the other hand, if  $\delta \phi = \mathcal{O}(1)$  and $\rho \lesssim  \delta \phi/{\sqrt{V_B}}$ near the maximum of the potential then for any regular solution at most $\phi$ can spend a time of order $V_B^{-1/2}$ near the maximum since $\rho$ would go to zero before one could spend much more time than $\mathcal{O}(V_B^{-1/2})$ near the maximum.  If $\delta \phi \ll 1$ and $\rho \lesssim  \delta \phi/{\sqrt{V_B}}$ then 
\be
\rho^2 V(\phi) \sim \rho^2 V_B \ll 1
\ee
and
\be
\rho^2 \phi'^2 \leq \rho^2 \Delta V_0 \lesssim \rho^2 V_B \ll 1
\ee
again assuming $V_T \gtrsim -V_B$ as before, and hence
\be \label{linreg}
\rho'^2 \approx 1
\ee
Then the solutions to the field equation under these circumstances is easy found in terms of Bessel functions to be
\be
\phi - \phi_B \approx \rho^{1-d/2}\Big[C_0 J_{d/2-1} (\sqrt{-V''(\phi_B)} \rho)+C_1 Y_{d/2-1} (\sqrt{-V''(\phi_B)} \rho) \Big]
\ee
where $\phi_B$ is the location of the barrier maximum and $C_0$ and $C_1$ are appropriate constants.   Then in a time interval
\be
\Delta \tau_3 \lesssim \frac{1}{\sqrt{-V''(\phi_B)} } \lesssim \sqrt{\frac{\delta \phi \Delta \phi_i}{\Delta V_i}} \lesssim \frac{\Delta \phi_i}{\sqrt{V_B}}
\ee
$\phi$ must cross the barrier or one will reach a zero of $\phi'$.

With the above comments, one is in a position to estimate the amount $E_0$ changes as it crosses the maximum region.  Noting that the field equation in terms of $E_0$ may be written as
\be
\partial_{\tau} E_0 = (d-1) \frac{\rho'}{\rho} \phi'^2
\ee
and so the change in energy as $\phi$ crosses the maximum will be
\be \label{deltaE}
\Delta E = (d-1) \int dt \frac{\rho'}{\rho} \phi'^2 \sim \frac{\Delta \tau_i}{\bar{\rho}} \bar{\rho}' V_B
\ee
where $\bar{\rho}$ and $\bar{\rho}'$ are the averages of $\rho$ and $\rho'$ over this region and $\Delta \tau_i$ either $\Delta \tau_2$ or $\Delta \tau_3$ depending upon the size of $\rho$ as $\phi$ crosses the maxima.   If $\rho \lesssim \delta \phi/\sqrt{V_B}$, as discussed before, the change in $E_0$ will generically be significant but if $\rho \gg \delta \phi/\sqrt{V_B}$ in this region then, presuming $\rho'^2 \lesssim 1$ (\ref{rhomaxbd}), 
\be
\Delta E \ll V_B
\ee
On the other hand, $\Delta E$ can be comparable to the minimum potential.  In particular, recalling that if the false vacuum potential $V_F$ is positive, $\rho \lesssim V_F^{-1/2}$ (\ref{rhomaxbd}), then
\be
\Delta E \gtrsim V_F \delta \phi \sqrt{\frac{V_B}{V_F}} \bar{\rho'}
\ee
and if the barrier height is large compared to the false potential ($V_B \gg V_F$) unless the width of the maxima region ($\delta \phi$) is either very small or the maximum of $\rho$ is quite close to the maximum of the potential ($\vert \bar{\rho'} \vert \ll 1$) then $\Delta E$ can be substantial compared to $V_F$. 

Combining all the above observations together, then under the above assumptions the amount of time $\phi$ will spend away from the minima is of order
\be
\Delta \tau  = \frac{\Delta \phi}{\sqrt{V_B}}
\ee
and the change in $\rho$ as the instanton travels through the wall will be at most of order $\Delta \tau$.   If the instanton starts near one the potential minima and exits that region only when $\rho \gg \Delta \tau$ then the change in $\rho$ through the wall is relatively small and the thin-wall approximation will be reliable.  Note under such circumstances the instanton necessarily ends close to the other minima, presuming the separation between the minima in $\phi$ is at most of order one, for $\rho$ must still either decrease to zero (in the asymptotically de Sitter case) or go to infinity (for the other asymptotics).   In the event of a minima that is infinitely far in $\phi$ away (e.g. a decompactification direction), under the above conditions $\phi$ will travel in a time $\Delta \tau$ to a point where $\vert V'(\phi) \vert$ is small compared to its average value and get to a region where one may, generically\footnote{As a matter of principle, one might be able to design potentials that falloff sufficiently slowly as they approach an asymptotic minima that, even in regions where $\vert V'(\phi) \vert$ is relatively small, $V(\phi)$ is still not close to its asymptotic value until $\phi$ is very large.  Such very flat large regions, however, would seem to make such potentials very highly tuned and their stability doubtful and I am not aware of any physically relevant potentials of this type.  Typically, the potentials of physical interest with asymptotic minima (e.g. the KKLT potentials \cite{KKLT}) falloff rather quickly, frequently exponentially, and one does not encounter such concerns.}, reasonably approximate the potential by its value at the asymptotic minima.  Further, for such large $\rho$ away from the extrema of the potential the energy $E_0$ (\ref{Edef}) will be approximately conserved.

\subsection{Thin barriers and thin-wall instantons}

Let us first consider the case where the barrier is thin ($\Delta \phi \ll 1$).    Note that, provided $V_T \gtrsim -V_B$, in view of the bound on $\phi'^2$ (\ref{philim})  the Einstein equation (\ref{Ein2}) impies $\rho''/\rho \lesssim \mathcal{O}(V_B)$ and the time scale for $\rho$ to go from zero, reach a maximum, and go to zero again (in the asymptotically de Sitter case) or go from zero to a large value (in the asymptotically flat or asymptotically AdS cases) is at least of order $V_B^{-1/2}$.   On the other hand, from the previous subsection, $\phi$ will transverse the region between minima in a time scale of order $\Delta \phi/\sqrt{V_B}$ so, if $\Delta \phi \ll 1$, $\phi$ must spend a relatively large amount of time close to at least one of the minima. 

Further, if the instanton is not well described by the thin-wall approximation then there is some point not near the minima where 
\be \label{rhosmall}
\rho \lesssim \Delta \tau = \frac{\Delta \phi}{\sqrt{V_B}}
\ee
As argued above, since the change in $\rho$ as one crosses the barrier is at most of order $\Delta \tau$, then (\ref{rhosmall}) must be obeyed away from the minima for any non-thin wall instanton.   Given the bound, $\phi'^2 \leq  2 \Delta V_0 \lesssim V_B$ then via the constraint (\ref{Ein1}) then necessarily
\be \label{rholinbd}
\rho'^2 = 1 + \mathcal{O}((\Delta \phi)^2) \approx 1
\ee
away from the minima of the potential.  Recalling that, in all cases, the instanton begins on the true vacuum side of the barrier with  $\rho' = 1$ when $\rho = 0$, then $\rho' \approx 1$ if and until $\rho \gg \Delta \tau$.   Then $\rho' \approx 1$ for any non-thin-wall (thin barrier) instanton unless and until $\phi$ approaches the false vacuum  except, potentially, for instantons which start sufficiently close enough to a positive true vacuum ($V_T >  0$) that $\rho$ passes through a maximum (at $\rho \sim V_T^{-1/2}$) and decreases to $\rho \lesssim \Delta \tau$ before leaving the neighborhood of the true vacuum.   

In fact this later scenario, where an instanton starts near the true vacuum and leaves this vicinity only when $\rho'$ is substantially negative (in the above case $\rho' \approx -1$ ), only yields singular instantons.  Recall that the field equation (\ref{Feqn})  in terms of the energy $E_0$
\be
E_0 = V(\phi) - \frac{\phi'^2}{2}
\ee
is
\be
E_0' = (d-1) \frac{\rho'}{\rho} \phi'^2
\ee
Remember, as well, the previous discussion below (\ref{philim}) indicating once $\rho'$ becomes negative it stays negative and in particular is either monotonically decreasing or bounded above by $-1$.  Then in such a scenario $E_0 \approx V_T$ (\ref{Ebds}) until $\phi$ leaves the neighborhood of the true vacuum and then $E_0$ monotonically decreases.  This is disastrous since for any regular instanton $\phi$ must ultimately come to rest, either at the second zero of $\rho$ in the asymptotically de Sitter case or asymptotically in the asymptotically flat or AdS cases, and at that point 
\be \label{E0bdy}
E_0 \rightarrow V(\phi)
\ee
However, for $E_0 < V_T$, as occurs in the above scenario, (\ref{E0bdy}) is impossible and $\phi$ never comes to rest, with the result the instanton becomes singular.

Hence $\rho' \approx 1$ for any non-thin-wall instantons starting on the true vacuum side of the potential barrier unless and until $\phi$ reaches the vicinity of the false vacuum.  Note the instanton necessarily ends near the false vacuum since $\phi$ spends at most time $\Delta \tau$ near the true vacuum and, as discussed above, $\phi$ must spend a time at least of order $V_B^{-1/2}$ near a minima for thin barriers.   In somewhat greater detail, in the asymptotically de Sitter case $\rho'$ must evolve from $+1$ to $-1$ and this can only happen if $\phi$ spends a relatively long time in a neighborhood of the false vacuum.  In the asymptotically flat or AdS context, $\rho$ must ultimately become large but if the instanton is not thin-wall after $\rho$ has crossed the barrier $\rho \sim \Delta \tau$ and $\rho$ is still quite small (e.g. compared to any AdS scale).

Viewed in time reverse, then any non-thin wall instanton starts close to the false vacuum, leaves the neighborhood of the false vacuum with $\rho \sim \Delta \tau$ and $\rho \approx - 1$ until the instanton ends somewhere on the true vacuum side of the barrier.   Until it leaves the vicinity of the false vacuum, $E_0 \lessapprox V_F$ (\ref{Ebds}) and then $E_0$ monotonically decreases until the instanton ends on the true vacuum side of the barrier at $\rho = 0$.   Since away from the minima $\rho \sim \Delta \tau$ and recalling that $\phi'^2 \sim V_B$ for a substantial fraction of the time $\phi$ spends between the minima, the energy lost crossing the barrier will be
\be \label{deltaE2}
\Delta E = (d-1) \int d\tau \frac{\rho'}{\rho} \phi'^2 \sim -V_B
\ee
If the barrier is too high and this energy decrease becomes too large then $\phi$ can never come to rest the and instanton is necessarily singular.  In particular if
\be \label{barriercrtierion}
V_B \gg V_F - V_T
\ee
then the energy on the true vacuum side of the barrier will be
\be
E_0 \lessapprox V_F -V_B \ll V_T
\ee
and $\phi$ can not come to rest near the true vacuum, let alone at any potential above $V_T$ and when $\rho$ goes to zero $\phi'^2 > 0$ and the instanton is necessarily singular.  In contrast, while  the energy of thin-wall instantons will generically change as $\phi$ crosses the barrier, typically by approximately $V_F - V_T$ (such instantons necessarily begin and end near the minima), $\rho$ away from the minima is large compared to $\Delta \tau$ and the change in $E_0$ small compared to (\ref{deltaE2}).  

As a result, as long as the barrier is thin ($\Delta \phi \ll 1$) the only regular instantons for potentials with relatively high barriers or nearly degenerate minima are well-described by the thin-wall approximation.  In the case of very negative minima ($V_T \lesssim -V_B$) the above considerations do not appear to provide any obstruction to non-thin-wall instantons, but as long as one is willing to assume that $\vert V_T \vert \ll V_B$ if $V_T < 0$ then (\ref{barriercrtierion}) is not just a sufficient but a necessary condition for thin barrier, thin-wall instantons.   The only way, under such conditions, for (\ref{barriercrtierion}) to be violated is if $V_B \sim V_F \neq 0$ which, since $\Delta \phi \ll 1$, implies the barrier tension will be small and unless the minima are nearly degenerate $\rho \sim \Delta \tau$ away from the minima.  This can be seen below in detail using the thin-wall approximation but also follows from the simple argument regarding bubble walls and energy reviewed in the introduction.  In particular, as the barrier tension becomes sufficiently small (as $\Delta \phi \rightarrow 0$ if $V_B \sim V_F$) as long as the minima are not nearly degenerate there are bubbles small compared to the scale set by $V_F$ and yet large compared to the size $\rho_0$ (\ref{smallbubblecrit}) needed to produce an approximately zero energy solution and so the use of the approximation is justified.

\subsection{Thin-wall instantons and non-thin barriers}

As the width of the barrier increases, generically the curvature of the potential near the maximum decreases and eventually one runs into the bound discussed in the last section (\ref{Vbd}).   As one might expect, then as one increases the width of the curve generically the initial value of $\phi$ ($\phi(0) = \phi_0$) moves towards the maxima, approaching the solution (\ref{simplesol}) in the limit the bound is saturated \cite{SteinhardtJensenI}.  It is, perhaps, not entirely obvious that one need not be terribly close to this limit for $\phi_0$ to move far away from the minima and once one does so $\phi_0$ can be remarkably independent of other features of the potential.   To illustrate this point, consider the generic quadratic potential
\be \label{simplepot}
V(\phi) = \frac{A_0 \phi^2}{12} \Big(3 \phi^2 - 4 (1+\lambda) \phi_2 \phi + 6 \lambda \phi_2^2 \Big) + V_0
\ee
where $A_0$, $\phi_2$, $\lambda$, and $V_0$ are constants and the perhaps awkward seeming parametrization is chosen so that
\be
V'(\phi) = A_0 \phi (\phi - \phi_1) (\phi - \phi_2)
\ee
where $\phi_1  =  \lambda \phi_2$ and so, for $A_0 > 0$ and $0 < \lambda < 1$, one has a potential with minima at $\phi = 0$ and $\phi = \phi_2$ and a relative maximum at $\phi_1$.  One can easily solve (\ref{simplepot}) for $A_0$ in terms of the height of the barrier ($V_1 = V(\lambda \phi_2)$) and solving for $\lambda$ in terms of the potential at the second minima ($V_2 = V(\phi_2)$) yields a quartic equation with roots  $\lambda(V_0, V_1, V_2) $ easily found by one's favorite symbolic manipulation program.  Doing so then one can then explicitly choose the height of the two minima ($V_0$ and $V_2$), the height of the barrier ($V_1$) and the width of the barrier ($\phi_2$) to be any desired values.  For this class of potentials one finds that the bound on the curvature near the maxima (\ref{Vbd}) means
\be
\phi_2^2 < \phi_c^2 \equiv \frac{ 12 (d - 1) (d -2)}{d} \frac{(1 - \lambda)}{ \lambda^2 (2 - \lambda)} \Big(1 - \frac{V_0}{V_1} \Big)
\ee
Generically $\phi_c$ is of order one as expected: $\lambda \rightarrow 0$ corresponds to $V_1 \rightarrow V_0$, $\lambda \rightarrow 1$ corresponds to $V_1 \rightarrow V_2$ and $\lambda \sim 1/2$ if $V_1 \gg V_0, V_2$.  

Given a definite potential (e.g. \ref{simplepot}) one can then numerically find the instanton using [(\ref{Ein2}), (\ref{Feqn})] and a simple power series expansion around $\tau = 0$ (where $\rho = \phi' = 0$), or in the case $\phi_0$ is very near the minima (\ref{phinear1}) or (\ref{phinear2}), to avoid numerical errors.  Given any specific value of $\phi(0) = \phi_0$ between the minima and maxima (without loss of generality, one may take $0 < \phi_0 < \lambda \phi_2$ for (\ref{simplepot})) one finds a unique instanton, albeit one which is generically singular.  For the sake of definiteness, let us consider tunnelling between two de Sitter minima.   If $\phi_0$ is too close to the maximum then $\phi'$ will encounter a zero while $\rho$ is still positive (presuming one enforced (\ref{Vbd})) while if it is too close to the minima $\rho$ will go to zero while $\phi' > 0$--typically in the later case for potentials such as (\ref{simplepot}) $\phi$ passes the second minima at $\phi_2$ and goes off to positive infinity as $\rho \rightarrow 0$.   The value of $\phi_0$ between these two regimes corresponds to the desired regular instanton.   Choosing an example potential where the barrier is both high compared to the minima and the minima are nearly degenerate with $\phi_2 = 9 \phi_c/10$ one obtains Figure 1 illustrating a variety of instantons starting midway between the minima and maxima.

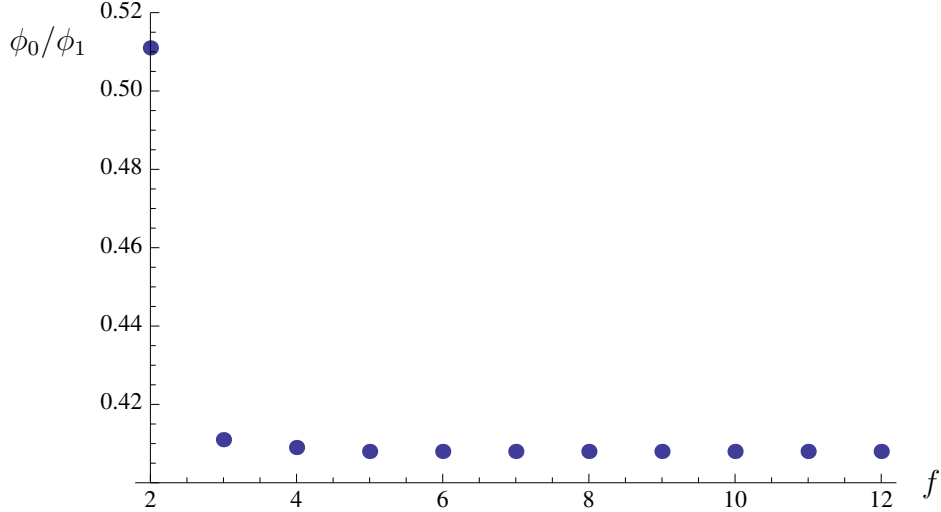
\begin{figure}[t]
\begin{picture} (0,0)
    	\put(-185,-20){$\phi_0/\phi_1$}
	         \put(160, -187){$f$}
    \end{picture}
    \centering

	\includegraphics[scale= 1.25]{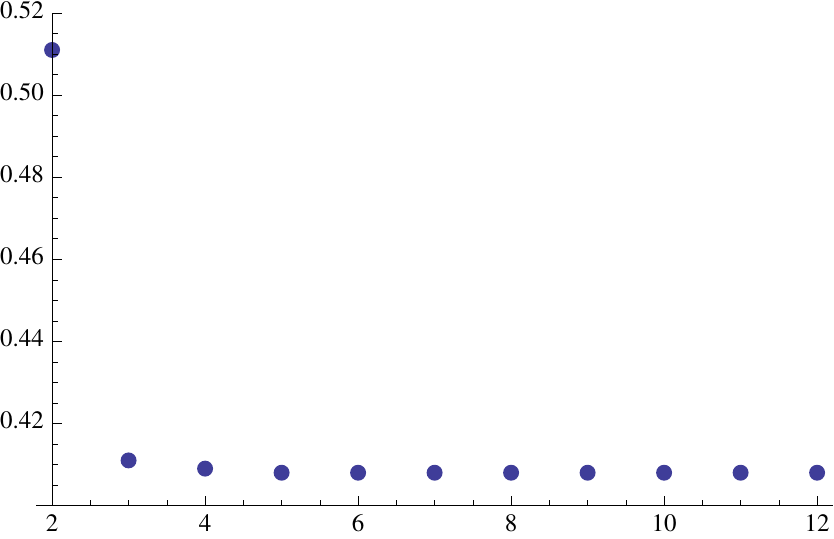}
	\caption{$\phi_0$ for potential (\ref{simplepot}) (in Planck units) for $d = 4$ with false vacuum potential $V_0 = 10^{-f}$, barrier height $V_1 = 1/10$, true vacuum potential $V_2 = 99 \, V_0/100$, potential width $\phi_2 = 9 \, \phi_c/10\approx 4.18$ and location of maxima $\phi_1 = \lambda \phi_2 \approx \phi_2/2$.}
	\label{ex1fig}
	\end{figure}
	
	In the light of such examples, it appears hard to imagine any criterion for thin-wall instantons as simple as that in the thin barrier case.  However, it is possible  to characterize, to some extent, the criterion necessary for a thin-wall instanton.  Given the results above, to obtain a thin-wall instanton it is necessary that $\phi$ starts near one of the relative minima and stays in that vicinity until $\rho \gg V_B^{-1/2}$.    That is until this time $V(\phi) \approx V_0$ and
\be
\rho \approx \frac{1}{\omega_0} \sin \omega_0 \tau
\ee
where
\be \label{omega0def2}
\omega_0 = \frac{V_0}{(d-1) (d-2)}
\ee
Then the statement that $\rho \gg V_B^{-1/2}$ while $\phi$ is still near a minima is equivalent to
\be
\sin \omega_0 \tau \gg \sqrt{\frac{V_0}{V_B}}
\ee
and consequently
\be
V_B \gg V_0
\ee
that is, a thin-wall instanton is possible for non-thin barriers only if the barrier is large compared to the potential at the minima (or large compared to the magnitude of the potential at the minima in the case of negative minima).   Hence I will only consider ``high'' barriers ($V_B \gg \vert V_F \vert, \vert V_T \vert$) for the remainder of this subsection.

Suppose that the instanton begins near a minima but $\rho \lesssim V_B^{-1/2}$ when $\phi$ enters the intermediate region so the change in $\rho$ as $\phi$ crosses the barrier is comparable to $\rho$.   Given the assumption of high barriers this implies $\rho' \approx \pm 1$ when $\phi$ leaves the vicinity of the minima.   If $\rho'$ is negative (i.e. $\phi$ has stayed near the minima as it has gone through a maxima and started decreasing) then, as discussed in the previous subsection, the energy $E_0$ decreases by order $V_B$ as $\phi$ travels through the intermediate region and $\phi$ can never come to rest.   Then, provided the instanton starts near a minima at $\tau = 0$, it will be well described as thin-wall if and only if when
\be
\frac{1}{\sqrt{V_B}} \ll \tau \ll \frac{1}{\sqrt{V_0}}
\ee
(where $\rho \approx \tau$) $\phi$ is still near the minima.  Specifically assuming that $\phi$ is sufficiently small that
\be
V'(\phi) \approx V'(\phi_0) + V''(\phi_0) (\phi - \phi_0)
\ee
where $\phi(0) = \phi_0$ then the field equation (\ref{Feqn}) becomes 
\be \label{Besseleq1}
\phi''+\frac{(d-1)}{\tau} \phi' \approx V'(\phi_0) + V''(\phi_0) (\phi - \phi_0)
\ee
which may be solved in terms of Bessel functions.  The only solution regular at $\tau = 0$ is
\be \label{Besselsol}
\phi \approx \phi_0 + \frac{V'(\phi_0)}{V''(\phi_0)} \Big[ \Gamma\Big(\frac{d}{2} \Big) 2^{d/2-1} (\lambda \tau)^{1-d/2} I_{d/2-1} (\lambda \tau) - 1 \Big]
\ee
where
\be
\lambda = \sqrt{V''(\phi_0)}
\ee

Then the instanton will be reliably thin wall if there are value of $\tau \sim \rho \gg V_B^{-1/2}$ where $\phi$ is still close to the minima--to be precise it is still true for such $\tau$ that $V(\phi) \approx V_0$ and $V'(\phi) = V'(\phi_0) + V''(\phi_0) (\phi - \phi_0)$.   Since (\ref{Besselsol}) is monotonically increasing, if $\phi$ starts close to the minima value, $\phi_0 \approx \phi_{min}$ where $V'(\phi_{min}) = 0$, then
\be
\frac{V'(\phi_0)}{V''(\phi_0)} \approx \phi_0 - \phi_{min}
\ee
and so this is simply a question of if $\phi_0$ begins close enough to $\phi_{min}$.  
For generic potentials one can estimate
\be
V''(\phi_0) \sim \frac{\Delta V_i}{(\Delta \phi_i)^2} \sim V_B
\ee
although one can also write potentials where the potential is highly curved (relatively speaking) near the minima and hence $V''(\phi_0) \gg V_B$.  
Since in this context
\be
\lambda \tau \approx \lambda \rho = \sqrt{\frac{V''(\phi_0)}{V_B}} \sqrt{V_B} \rho
\ee
for generic potentials ($V''(\phi_0) \sim V_B$)  $\rho \lesssim V_B^{-1/2}$ implies $\lambda \tau \lesssim 1$.  Then for generic potentials as long as $\phi$ starts close to the minima it will remain so until $\rho \gg V_B^{-1/2}$ and the instanton will be well described as thin-wall.    If $V''(\phi_0) \gg V_B$, then $\lambda \tau \gg 1$ even for $\rho \sim V_B^{-1/2}$ and so (\ref{Besselsol}) becomes
\be
\phi \approx \phi_0 + \frac{V'(\phi_0)}{V''(\phi_0)} \frac{\Gamma \Big(\frac{d}{2} \Big) 2^{d/2-1}}{\sqrt{2 \pi}} (\lambda \tau)^{(1-d)/2} e^{\lambda \tau} \Big[1+ \mathcal{O}\Big(\frac{1}{\lambda \tau}\Big) \Big]
\ee
and $\phi_0$ must be exponentially close to the minima if the instanton is to be well described as thin-wall.   It is also possible to use (\ref{phinear1}) or (\ref{phinear2}) to perform an analogous analysis to the above--namely making sure that $\phi$ remains sufficiently small until $\rho$ is large enough to make the thin-wall approximation valid--without approximating $\sin\omega_0 \tau \approx \omega_0 \tau$.  This may be useful in dealing numerically with concerns one might have for potentials barriers which are high but not very high, although without approximating $\rho$ as linear it seems difficult to make as clear analytic statements as above.  

	Perhaps the most prominent example of a potential which does not obviously correspond to a thin-wall instanton is the KKLT potential \cite{KKLT} (see Figure 2)
	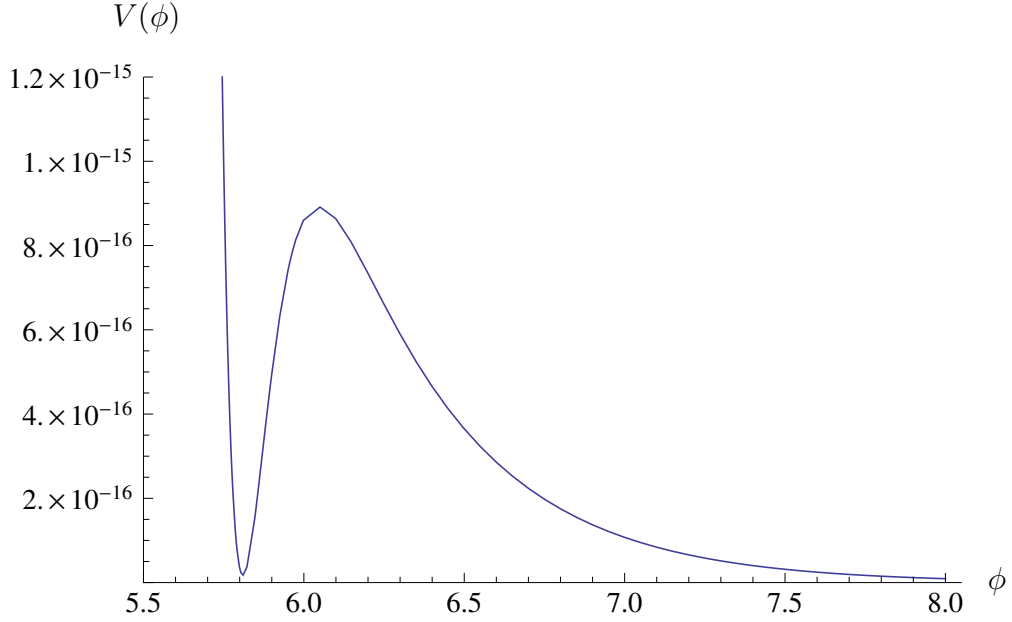
\begin{figure}[t]
\begin{picture} (0,0)
\put(-141,10){$V(\phi)$}
	         \put(190, -203){$\phi$}
    \end{picture}
    \centering

	\includegraphics[scale= 1.5]{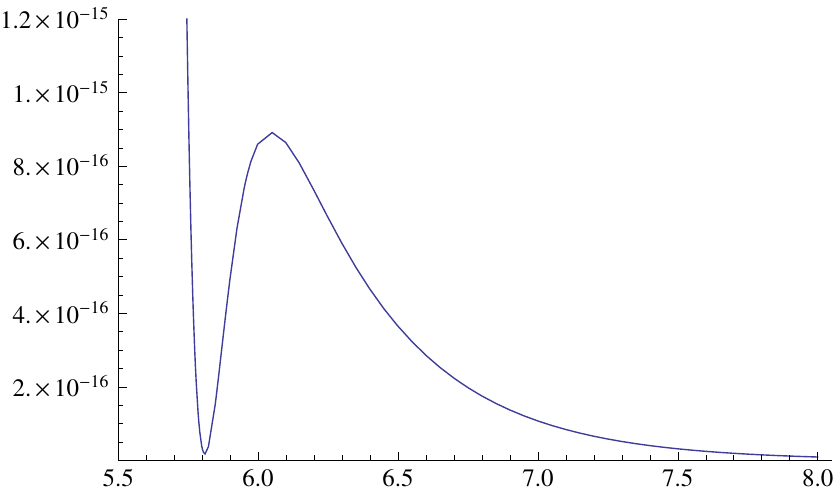}
	\caption{KKLT  potential (\ref{KKLTpot}) (in Planck units) for $W_0 = -10^{-4}$, $A = 1$, $a = 0.1$, and $D = 3 \times 10^{-9}$ versus canonically normalized scalar $\phi$}
	\label{KKLT1fig}
	\end{figure}
	
	\be \label{KKLTpot}
	V(\phi) =  \frac{a A e^{-a \sigma}}{2 \sigma^2} \Big( \frac{1}{3} \sigma a A e^{-a \sigma} + W_0 + A e^{-a \sigma} \Big) + \frac{D}{\sigma^3}
	\ee
	with
	\be
	\sigma = e^{\sqrt{2/3} \,  \, \phi}
	\ee
	yielding, at least for the parameters focused on in \cite{KKLT} and given in the caption in Figure 2, a barrier which is relatively high ($V_B/V_F \approx 50.47$) but not overly thin.\footnote{The width of potentials which only asymptotically approach a minima (e.g. $V(\phi) = 0$) can be defined by finding the $\phi$ at which $V'(\phi)$ becomes small compared to the average value of $V'(\phi)$ on a given side of the barrier.  Provided this width is not large in Planck units, one can use the above results without modification.}  Following the procedure outlined above, one quickly finds $\phi'$ changes sign while $\rho$ remains positive unless $\phi$ starts very close to the false vacuum.  In fact, for this potential it seems difficult to find a $\phi_0$ sufficiently close to $\phi_F$, the location of the false vacuum, that $\phi$ becomes monotonic before one runs into numbers so small one may doubt (at least without the assurance of a numerical expert) whether the calculation is truly under control.    However, if one shows that choosing $\phi_0 - \phi_F$ sufficiently small that $\rho$ reaches values large compared to $V_B^{-1/2}$ while $\phi$ is still close to $\phi_F$ (in the sense that the above approximations regarding $V(\phi)$ and $V'(\phi)$ are still reliable) and yet the resulting instanton still results in $\phi'$ reversing signs before $\rho$ can go to zero (i.e. the regular solution requires an even smaller value of $\phi_0 - \phi_F$) then one will have ruled out non-thin-wall instantons.  In particular for potentials such as (\ref{KKLTpot}) which might be described as having a high, but not terribly high, barrier one might be inclined to try to prove the stronger criterion that $\rho'$ is decreasing before $\phi$ leaves the vicinity of $\phi_F$ by demanding that $\phi_0 -\phi_F$ is sufficiently small that $\rho$ can reach its maximum possible value  ($\rho \approx \omega_0^{-1}$ where $\omega_0$ is, as before, given by (\ref{omega0def2}) and $V_0 \approx V_F$) before $\phi$ can get far away from $\phi_F$.   Then, as argued before, one is guaranteed the only regular solutions must be well-described as thin-wall.  Undertaking this second objective for (\ref{KKLTpot}) with the given parameters, after a bit of trial and error one finds $\phi_0 - \phi_F = 10^{-220}$ still is too large a value (i.e. the resulting instanton still results in $\phi'$ encountering a zero while $\rho$ is positive) and yet even by the time $\rho$ reaches such a maximum value ($\rho \approx \omega_0^{-1}$) one finds $\phi - \phi_F \approx 1.45 \times 10^{-10}$,
	\be
	V(\phi) \approx V_F \Big(1+ 1.71 \times 10^{-16} \Big)
	\ee
	 and 
	 \be
	 V'(\phi) \approx \Big[V'(\phi_0) + V''(\phi_0) (\phi - \phi_0)\Big] \Big(1+ 2.21 \times 10^{-9}\Big)
	 \ee
 In other words, the regular instanton for the given potential requires $\rho'$ to be negative before it leaves the vicinity of the false vacuum and will be well described by the thin-wall approximation.

\subsection{Misconceptions regarding the thin-wall approximation}

In the discussion of the thin-wall approximation, one occasionally encounters assertions that the approximation implies one may simply dropping the friction term in the field equation (\ref{Feqn}) or that the ``wall'' may be treated as infinitesimally thin compared to any other scales in the problem.   Both of these statements are false and I will end this section with a discussion of that fact.   While as previously discussed in the thin-wall approximation away from the extrema of the potential one will have an approximately conserved energy $E_0$ (\ref{Edef}), it is not consistent to simply drop the friction term $ \frac{\rho'}{\rho} \phi' $ from the field equation (\ref{Feqn})
\be \label{Feqn2}
\phi'' + (d - 1) \frac{\rho'}{\rho} \phi' = V'(\phi)
\ee
for all times.  In particular, for any regular solution expanding the Einstein and field equations (\ref{Ein1})-(\ref{Feqn}) near $\tau = 0$
\begin{eqnarray}
\phi &=& \phi_0 + \frac{V'(\phi_0)}{2 d} \tau^2 + \mathcal{O}(\tau^4) \nonumber \\
\rho &=& \tau - \frac{V(\phi_0)}{6 (d - 1) (d - 2)} \tau^3 + \mathcal{O}(\tau^5)
\end{eqnarray}
Then
\be
\phi'' = \frac{V'(\phi_0)}{d} + \mathcal{O}(\tau^2)
\ee
and
\be
(d - 1) \frac{\rho'}{\rho} \phi' = (d-1) \frac{V'(\phi_0)}{d} + \mathcal{O}(\tau^2)
\ee
and the friction term is comparable to the other terms in (\ref{Feqn2}).  

More generically, one can rewrite the Einstein and field equations in terms of the potential and the energy
\be
E_0 = V(\phi) - \frac{1}{2} {\phi'}^2
\ee
Noting the field equation (\ref{Feqn}) is equivalent to
\be \label{E0prime}
E_0' = (d-1) \frac{\rho'}{\rho} \phi'^2
\ee
then a few lines of algebra yield
\be
\rho = \Big[ \frac{E_0}{(d - 1) (d - 2)} + \frac{{E_0'}^2}{4 (d - 1)^2 (V - E_0)^2} \Big]^{-1/2}
\ee
and
\be \label{Eevol}
E_0'' = \frac{\sqrt{2} E_0'}{\sqrt{V-E_0}}-\frac{ (2 d - 1) {E_0'}^2}{2 (d - 1) (V - E_0)} - \frac{2 (V - E_0)}{d-2} \Big[ (d-1) (V - E_0) + E_0 \Big]
\ee
Then if one took $E_0$ to be exactly constant (i.e. simply dropped the friction term in (\ref{Feqn2}))  in (\ref{Eevol}) one would be forced to conclude that $V(\phi)$ is exactly constant as well (e.g. $V(\phi) = E_0$ in the case of non-negative potentials).  This does not imply any contradiction with the above results; if $\rho$ is sufficiently large away from the extrema of the potential that one has an approximately conserved $E_0$, $\phi$ transverses these regions relatively quickly and a nonzero $E_0'$ or $E_0''$ has a small effect on $E_0$ in that time.

It is sometimes asserted that in the thin-wall approximation one may simply regard the wall as infinitely thin and hence one needs only satisfy the junction conditions at the wall to find an appropriate instanton and evaluate the on-shell action.  In addition to the generic problems described in the introduction that this produces, for generic potentials, as discussed above in detail,  one can argue only a single regular instanton (with $\phi$ monotonic) exists \cite{SteinhardtJensenI} corresponding to starting $\phi$ at some finely-tuned initial value $\phi_0$ and $\phi$ takes a finite amount of time ($\Delta \tau \sim \Delta \phi/{\sqrt{V_B}}$ for the potentials discussed) to cross the barrier and hence there is some finite thickness wall.  More generally, if there were a truly wall infinitesimally smaller than any other scales  in the problem (and in particular the friction term in the equation for $\phi$), then $E_0$ would be conserved (\ref{E0prime}) as one crosses the wall.  Hence, after crossing the wall $\phi'$ becomes nonzero
\be \label{phidif}
\frac{\phi'^2}{2} \approx V_F - V_T
\ee
and since as long as $V'(\phi)$ is truly ignored then in this minima
\be
\phi' = \frac{C_2}{\rho^{d-1}}
\ee
for a nonzero constant $C_2$ (easily obtained from (\ref{phidif})) and at least in the asymptotically de Sitter case $\phi'$ diverges as $\rho \rightarrow 0$ and the instanton is singular.  

In fact, one can make a more elegant argument showing this result with rather weaker assumptions using (\ref{consistcondit}), the condition required if $\phi'$ is to vanish at the second zero of $\rho$.  To see this, let us divide up the instanton into a region between $0$ and $\tau_0$ where $V(\phi) \approx V_0$ for some constant $V_0$, a region from $\tau_1$ to $\tau_f$ where $V(\phi) \approx V_2$ for some constant $V_2$ and a wall region between $\tau_0$ and $\tau_1$.  From (\ref{consistcondit})
\begin{eqnarray}
0 &=&  \int_0^{\tau_f} ds V(\phi) \rho^{2 d - 3} \rho' \nonumber \\
&=& \underbrace{\int_0^{\tau_0} ds V(\phi) \rho^{2 d - 3} \rho'}_{\approx \frac{V_0}{2 (d-1)}  \rho^{2 (d-1)}(\tau_0)}+  \int_{\tau_0}^{\tau_1} ds V(\phi) \rho^{2 d - 3} \rho' + \underbrace{\int_{\tau_1}^{\tau_f} ds V(\phi) \rho^{2 d - 3} \rho'}_{\approx -\frac{V_2}{2 (d-1)}  \rho^{2 (d-1)}(\tau_1)} \nonumber \\
&\approx&   \frac{V_0}{2 (d-1)}  \rho^{2 (d-1)}(\tau_0)-\frac{V_2}{2 (d-1)}  \rho^{2 (d-1)}(\tau_1) + \frac{V(\phi)}{2 (d - 1)} \rho^{2 d - 2} \Bigg \vert_{\tau_0}^{\tau_1} \nonumber \\
& -&  \frac{1}{2 (d - 1)} \int_{\tau_0}^{\tau_1} ds V'(\phi) \phi'  \rho^{2 d - 2} \nonumber \\
&=& -  \frac{1}{2 (d - 1)} \int_{\tau_0}^{\tau_1} ds \, V'(\phi) \phi'  \rho^{2 d - 2}
\end{eqnarray}
Note the approximations in the above can be made arbitrarily good by choosing $\tau_0$ and $\tau_1$ appropriately.  The junction conditions require $\rho$ is continuous between the two regions (although its derivative need not be) and hence as $\tau_1 \rightarrow \tau_0$, $\rho$ becomes arbitrarily close to a  constant for $\tau_0 \leq \tau \leq \tau_1$ and one would conclude
\be
0 \approx  \int_{\tau_0}^{\tau_1} ds \, V'(\phi) \phi'= V_2 - V_0
\ee
in contradiction of the desired assumption.  That is, as long as the vacua are not exactly degenerate $\phi$ acquires a nonzero velocity as it crosses the wall and this can not be dissipated fast enough to avoid making the instanton singular.

One might reasonably wonder how the above criterion
\be \label{wallconsist2}
0 \approx  \int_{\tau_0}^{\tau_1} ds \, V'(\phi) \phi'  \rho^{2 d - 2}
\ee
is consistent with the statements in the previous subsections and in particular why taking the limit of arbitrarily thin barriers does not contradict (\ref{wallconsist2}).  If one defines $\rho$ in the wall region by taking
\be
\rho = \rho_0 + \delta \rho
\ee
for some constant $\rho_0$ (e.g. the average of $\rho$ over the wall region) then (\ref{wallconsist2}) becomes
\be \label{wallconsist3}
0 \approx  \int_{\tau_0}^{\tau_1} ds \, V'(\phi) \phi'  \Big(1+\frac{\delta \rho}{\rho_0}\Big)^{2 d - 2}
\ee
If $\delta \rho \ll \rho_0$ then (\ref{wallconsist3}) becomes
\be \label{wallconsist4}
0 \approx V_2 - V_0 + 2 (d - 1)  \int_{\tau_0}^{\tau_1} ds  \frac{\delta \rho}{\rho_0} V'(\phi) \phi'
\ee
although in the case $\delta \rho \sim \rho_0$, (\ref{wallconsist4}) becomes an order of magnitude estimate.  Then
\be \label{deltarhobd}
\delta \rho \gtrsim \rho_0 \frac{\Delta V}{V_B}
\ee
where $\Delta V = V_2 - V_0$ and the inequality reflects possible internal cancelations in the integral in (\ref{wallconsist4}) (e.g. if $\delta \rho$ is approximately equal in regions on both sides of the barrier).   Then if one requires $\delta \rho \ll \rho_0$, $V_B \gg \Delta V$, consistent with the above results regarding the thin-wall approximation.  Using the earlier results that $\vert \delta \rho \vert  \lesssim \Delta \phi/{\sqrt{V_B}}$ then
\be
\rho_0 \lesssim \frac{1}{\sqrt{\vert \Delta V \vert}} \sqrt{\frac{V_B}{\vert \Delta V\vert }} \Delta \phi
\ee
If one takes sufficient thin barriers not only does $\delta \rho$ become small but so does $\rho_0$.  This is simply the fact alluded to in the introduction that small tension walls result in small bubbles.  Note, however, the above does not force $\delta \rho/{\rho_0}$ to become small for thin barriers, consistent with (\ref{deltarhobd}).  

\setcounter{equation}{0}

\section{Thin wall decays}

\subsection{General procedure for dS decays}

As previously mentioned, an $SO(d)$ symmetric instanton may be written as
\be
ds^2 = d\tau^2 + \rho^2(\tau) d\Omega_{d-1}
\ee
and $\phi(\tau)$ and the on-shell action may be written as (\ref{act2})
\be \label{onact2}
S_E = -\frac{2 \kappa \Omega_{d-1}}{d-2} \int_{0}^{\tau_f} d\tau \rho^{d-1} V(\phi)
\ee
where the tunnelling rate is given by
\be \label{dcyrate1}
\Gamma = A e^{-B}
\ee
where
\be
B = S_E  - S_b
\ee
where $S_b$ is the on-shell action of the original (e.g. false) vacuum.  

Choosing $\tau$ so that $\rho(0) = 0$ and at $\tau = 0$ the instanton is near the true vacuum one can split up the on shell action into three pieces
\be \label{act5}
S_E = -\frac{2 \kappa \Omega_{d-1}}{d-2}\Bigg[  \underbrace{\int_{0}^{\tau_0} d\tau \rho^{d-1} V(\phi)}_{I_0} + \underbrace{\int_{\tau_0}^{\tau_1} d\tau \rho^{d-1} V(\phi)}_{I_1} + \underbrace{\int_{\tau_1}^{\tau_f} d\tau \rho^{d-1} V(\phi)}_{I_2}  \Bigg]
\ee
where $I_0$ corresponds to the nearly true vacuum region with $V(\phi) \approx V_0$, $I_1$ the action for the wall where $\rho \approx \rho_0$ for some constant $\rho_0$, and $I_2$ for the nearly de Sitter region with $V(\phi) \approx V_2$.  For the remainder of this section, I will always make the above approximations.  Then for $0 \leq \tau \leq \tau_0$
\be
\rho= \frac{1}{\omega_0} \sin \omega_0 \tau
\ee
where
\be
\omega_0^2 = \frac{V_0}{(d-1) (d-2)}
\ee
and one takes the obvious limits to recover the decays to flat and asymptotically AdS vacuums, namely in the flat case for $0 \leq \tau \leq \tau_0$
\be
\rho =  \tau
\ee
and in the AdS case
\be
\rho = \frac{1}{\omega_0} \sinh \omega_0 \tau
\ee
for $0 \leq \tau \leq \tau_0$ where, after relabeling,
\be
\omega_0^2 = \frac{\vert V_0 \vert}{(d-1) (d-2)}
\ee
For asymptotically de Sitter decays for $\tau_1 \leq \tau \leq \tau_f$
\be
\rho= \frac{1}{\omega_2} \sin \omega_2 (\tau - \delta)
\ee
where
\be
\omega_2^2 = \frac{V_2}{(d-1) (d-2)}
\ee
as well as $\rho(\tau_f) = 0$, i.e.
\be
\omega_2 (\tau_f - \delta) = \pi
\ee
and the constant $\delta$ is chosen so that $\rho(\tau_0) = \rho(\tau_1) = \rho_0$.  This actually leads to two to two possible choices of $\delta$, namely $\delta_0$
\be
\omega_2 (\tau_1 - \delta_0) = \arcsin(\omega_2 \rho_0)
\ee
where $0 < \omega_2 (\tau_0 - \delta_0) < \pi/2$ or $\delta_1$
\be
\omega_2 (\tau_0 -\delta_1) = \pi - \arcsin(\omega_2 \rho_0)
\ee
and hence $\pi/2 < \omega_2 (\tau_0 - \delta_1) < \pi$.  For the sake of compactness it is useful to define a sign $s_0$ such that
\be
\delta = 
\begin{cases} \delta_0 & \text{$s_0 = 1$}\\
\delta_1 &\text{$s_0 = -1$}
\end{cases}
\ee
and so
\be
\rho'(\tau_1) = \cos \omega_2 (\tau_1 - \delta) = s_0 \sqrt{1 - \sin^2 \omega_2 (\tau_1 - \delta)} = s_0 \sqrt{1 -\omega_2^2 \rho_0^2}
\ee
As long as $V(\phi) \geq 0$ in the wall region then, due to (\ref{Ein2}), or if one prefers the junction conditions, one must choose $\delta$ consistent with
\be \label{deltarho}
\rho'(\tau_0) \geq \rho'(\tau_1)
\ee

$I_2$ can be written in general dimensions in terms of a hypergeometric function as
\be
I_2 = \frac{V_2}{\omega_2^d} \Bigg[s_0 \sqrt{1 -\omega_2^2 \rho_0^2} \, F\Big(\frac{1}{2}, 1 - \frac{d}{2},  \frac{3}{2}, 1 - \omega_2^2 \rho_0^2\Big) + \frac{\sqrt{\pi}}{2} \frac{\Gamma \Big(\frac{d}{2} \Big)}{\Gamma \Big( \frac{d+1}{2} \Big)} \Bigg]
\ee

On the other hand, to leading order $\rho(\tau_0) = \rho(\tau_1) = \rho_0$ is a constant in the wall region and so
\be
I_1 = \rho_0^{d-1} \int_{\tau_0}^{\tau_1} d\tau V(\phi) = \frac{V_2}{\omega_2^d} (\omega_2 \rho_0)^{d-1} V_a
\ee
where the wall tension
\be
V_a = \omega_2 \int_{\tau_0}^{\tau_1} d\tau \frac{V(\phi)}{V_2}  
\ee
is, to leading order, independent of $\rho_0$ as long as the thin wall approximation holds.  
Roughly speaking, using the result from the previous section that the time spent in the wall is of order $\Delta \phi/{\sqrt{V_B}}$ ,where $\Delta \phi$ is the width of the barrier and $V_B$ its height
\be
V_a \sim \Delta \phi \sqrt{\frac{V_B}{V_2}}
\ee
 More precisely, 
\be
V_a = \omega_2 \int_{\phi_1}^{\phi_2} \frac{d \phi}{\phi'} \frac{V(\phi)}{V_2}
\ee
where $\phi(\tau_0) = \phi_1$ and $\phi(\tau_1) = \phi_2$ and in the thin wall approximation $\phi(\tau)$ is determined at leading order just in terms of the potential; on each side of the barrier one has an approximately conserved energy
\be \label{E0def3}
E_0 = V(\phi) - \frac{\phi'^2}{2} 
\ee
and near the maximum (\ref{phimax})
\be
\frac{\phi'^2}{2} \approx V(\phi) - V(\phi(\tau_2)) + \phi'(\tau_2)^2 \approx V(\phi) - V_0 \sim V_B
\ee
where $\tau_2$ is some time chosen such that $\phi$ is on the true vacuum side of the barrier and  $V(\phi(\tau_2)) \sim V_B$ but $\phi$ is still sufficiently far away from the maximum that $E_0$ is still approximately conserved.  As discussed in the previous section, for generic potentials $E_0$ will not be conserved sufficiently close to the maximum and will change by an amount $\Delta E$ (\ref{deltaE2}) as $\phi$ crosses the top of the barrier but for specific potentials it may be true that $\Delta E$ is negligably small and $E_0$ is approximately conserved away from the minima and one may use (\ref{E0def3}) for the entire wall region.  In the decays to de Sitter or flat spaces I will assume $V_a > 0$, as will be automatic if one, as usual, takes potentials monotonically go from minima to the barrier, although in the decays to negative energy vacua I will also comment on the case $V_a < 0$.

Then calculating $I_0$ for each appropriate case one can then calculate the on-shell action.  The thin wall approximation, as well as the symmetry assumptions, has reduced the problem of finding the instanton action to a function of a single variable $\rho_0$, as well as parameters characterizing the potential.  Since any solution extremizes the action, by demanding that the derivative of $S_E$ with respect to $\rho_0$, or, depending upon the circumstances, a more convenient function of $\rho_0$, allows one to solve for $\rho_0$, just as in Coleman-de Luccia \cite{CdL}.  It is worth emphasizing, however, that the procedure of assuming the instanton is well-described by the thin-wall approximation and then checking $\rho_0$ is large compared to the thickness of the wall does not in any sense justify the use of the approximation.  The instantons described in Figure 1 would pass such a test, since if they were well described by the thin-wall approximation they would give rise to large tension walls and large bubble instantons, but they start well away from the minima and are not in any sense thin-wall.   On the other hand, if the thin-wall approximation is justified this criterion is automatic.

Given a thin-wall instanton, since we are ultimately interested in the decay rate one can just as well extremize $B$ or, pulling out some common parameters, the related dimensionless function $K_0$
\be
B = S_E - S_b = -\frac{2 \kappa \Omega_{d-1}}{d - 2} \frac{V_2}{\omega_2^d} K_0
\ee
Note then a negative $K_0$ corresponds to a decay rate that is exponentially suppressed while a positive $K_0$ one which is exponentially enhanced.

\subsection{Comparison with Coleman-de Luccia}

Given the divergence between the arguments made in the introduction and the previous results by Coleman and de Luccia \cite{CdL}, as well as the following work by Parke \cite{ParkeBubbles}, it is important to understand the differences between these calculations and the method described above.  The first, rather obvious, difference is a somewhat different form of the on-shell action used.  Rather than using (\ref{onact2}) one can, assuming the solution is $S0(d)$ symmetric, write out the curvature scalar $R$ explicitly and then perform an integration by parts
$$
S_E = -\kappa \Omega_{d-1} \int_{0}^{\tau_f} d\tau \rho^{d-1} \Big[ -2 (d-1) \frac{\rho''}{\rho} - \frac{(d-1) (d-2)}{\rho^2} (\rho'^2 - 1) - \frac{1}{2} \phi'^2 - V(\phi) \Big]
$$
\be  \label{Spartsact}
= \kappa \Omega_{d-1} \Bigg[ 2 ( d-1) \rho^{d-2} \rho' \Bigg\vert_{0}^{\tau_f} - \int_0^{\tau_f} d\tau \rho^{d-1} \Big[ (d-1) (d-2) \frac{(\rho'^2+1)}{\rho^2} -\frac{1}{2} \phi'^2 -V(\phi) \Big] \Bigg]
\ee
It is worth noting, in passing, that this action is unbounded from both above and below--the action can be made arbitrarily large and positive by making high frequency small amplitude oscillations in $\phi$ (as one would expect due to the connection between the action and energy in a context where the latter makes sense) but high frequency small amplitude oscillations for $\rho$ (taking the oscillations to be of compact support and hence not contributing to the boundary terms for the sake of simplicity) makes the action unbounded from below.  The latter is simply the famous statement \cite{GibbonsHawkingactionunbded} that the Euclidean gravitational action is unbounded from below due to oscillations in the conformal factor ($\rho$, once one makes an appropriate redefinition of $\tau$).  The finite action solutions are actually saddle points of the action.
Imposing the constraint (\ref{Ein1})
\be
\rho'^2 = 1 + \frac{\rho^2}{(d-1) (d-2)} \Big(\frac{1}{2} \phi'^2 - V(\phi) \Big)
\ee
 on (\ref{Spartsact}) gives
\be
S = -\kappa \Omega_{d-1} \Bigg[ - 2 ( d-1) \rho^{d-2} \rho' \Big \vert_{0}^{\tau_f} + \int _0^{\tau_f} d\tau \Big[ 2 (d-1) (d-2) \rho^{d-3} - 2 V(\phi) \rho^{d-1} \Big] \Bigg]
\ee
For asymptotically de Sitter instantons  $\rho(0) = \rho(\tau_f) = 0$ and for any regular instantons this boundary term will vanish and hence
\be 
S =  -\kappa \Omega_{d-1}  \int_0^{\tau_f} d\tau \Big[ 2 (d-1) (d-2) \rho^{d-3} - 2 V(\phi) \rho^{d-1} \Big] 
\ee
For asymptotically flat solutions where asymptotically
\be
\rho \rightarrow \tau
\ee
or the asymptotically anti-de Sitter case where
\be
\rho \rightarrow \frac{1}{\omega_2} \sinh \omega_2 \tau
\ee
one would have to specify a regularization scheme and deal with possible cutoff dependence and possible finite pieces left over if one wanted to rigorously justify simply dropping this surface term.  Especially for the asymptotically flat case it is somewhat difficult to see the virtue of this approach, since it has converted a convergent answer (\ref{onact2}) into one with two divergent pieces.  The asymptotically anti-de Sitter case, on the other hand, requires regularization either way.      Provided one is in an asymptotically de Sitter spacetime or for other asymtptotics there are regularizations schemes that justify dropping this boundary term then one has two different formulas for the on-shell action (\ref{act1}) and (\ref{act2}) and demanding that they be equal implies
\be \label{act6}
\int_0^{\tau_f} \rho^{d-1} V(\phi) = (d-2)^2 \int_0^{\tau_f} d\tau \rho^{d-3}
\ee
Note that for some instantons, namely decays between asymptotically AdS solutions where $V(\phi)$ is everywhere negative, (\ref{act6}) is clearly impossible and one must conclude either such solutions simply fail to exist or any regularization scheme must produce a nonzero contribution from this boundary term.

For tunnelling involving asymptotically de Sitter  solutions the above complications are avoided and the two different forms of the action truly are equivalent (for regular instantons).  The key difference, in the dS case, between the results I will describe and the previous work of Coleman-de Luccia and Parke  \cite{CdL, ParkeBubbles} is that earlier analysis implicitly neglected one of the effects of backreaction in evaluating the action and hence dropped a term which should have been retained.  Dividing up $B$ in the same fashion as (\ref{act5})  this previous work asserts that the part of $B$ outside the wall  (i.e. $\tau > \tau_1$), where the potential of the instanton matches (strictly speaking is extremely close to) the potential of the false vacuum background, vanishes identically.   The problem with this assertion is that while after $\tau_1$ the potentials between the two different solutions are approximately the same, the instanton generically does not last the same period of time as the background solution (due to the backreaction of $\phi$ on the metric (\ref{Ein2})) and hence the value of $\rho(\tau_1)$ does not match between the two solutions.  Said another way, if one started out both the background solution and tunnelling instanton in the false vacuum at the same time then one would be guaranteed $\rho$ at the wall in the tunneling solution would match the background solution but since there is a mismatch in these times there will be a mismatch in $\rho$.  Since $\rho$ is relatively large at the wall, by assumption, even a small difference in $\tau_f$ can result in a significant difference in $\rho(\tau_1)$.  One could avoid such a mismatch of $\rho$ by starting both instantons at $\rho(0) = 0$ in the false vacuums but one instanton will end, generically, before the other and one does not obtain the simple expressions given in \cite{CdL, ParkeBubbles}. 

At least in the asymptotically de Sitter case, and possibly more generally, one could as a matter of principle use the same form of the action
\be
S_E = -\kappa \Omega_{d-1} \int_0^{\tau_f} d\tau \Big[ 2 (d-1) (d-2) \rho^{d-3} - 2 V(\phi) \rho^{d-1} \Big]
\ee
and simply extremize $S_E$ without attempting to subtract off $S_b$, piece by piece, at an intermediate stage and avoid these complications.  However, it does not appear obvious as to how one could evaluate the portion of the wall action independent of $V(\phi)$
\be \label{wallint}
\int_{\tau_0}^{\tau_1} d \tau \rho^{d-3}
\ee
without knowing precisely the amount of time one has spent in the wall.  As discussed in the previous section, it is not generally consistent to claim (\ref{wallint}) is infinitesimally small.  One could use (\ref{act6}) to evaluate (\ref{wallint}) but one might as well simply evaluate the action (\ref{onact2}) directly as discussed above and done so explicitly in the next subsections.

\subsection{dS to flat decays}

Note that in this case since $\rho' = 1$ for $\tau \leq \tau_0$ and $\rho' = \cos \omega_2 (\tau - \delta)$ for $\tau_1 \leq \tau \leq \tau_f$ the change in $\rho'$ across the wall is automatically non-positive, as required (\ref{deltarho}).  The only remaining part of the action not calculated yet, $I_0$, is in this case arbitrarily small and may be neglected since in that regime $V(\phi) \approx 0$.  Defining a dimensionless variable
\be
x   = \omega_2 \rho_0
\ee
and using 
\be
K_0 =  V_a x^{d - 1} + s_0 \sqrt{1 - x^2} \, F(\frac{1}{2}, 1 -\frac{d}{2}, \frac{3}{2}, 1 - x^2) - \frac{\sqrt{\pi}}{2} \frac{\Gamma \Big(\frac{d}{2} \Big)}{\Gamma \Big(\frac{d+1}{2} \Big)}
\ee
and as a result
\be
\partial_ x K_0 = x^{d - 2} \Big[ (d - 1) V_a - \frac{s_0 x}{\sqrt{1-x^2}} \Big]
\ee
Hence $K_0$ has a nontrivial (i.e. $x \neq 0$) extrema if and only if
\be
s_0 = 1
\ee
at
\be
x_c = \frac{ (d-1) V_a}{\sqrt{1+(d-1)^2 V_a^2}}
\ee
Note that, as previously argued, for low tension walls $x_c$ becomes relatively small while as $V_a \rightarrow \infty$ it approach unity (i.e. $\rho_0$ is of order the false vacuum cosmological scale).

For $d =4$, on-shell $K_0$ adopts the relatively simple expression
\be \label{4dflatdec}
K_0(x_c, d = 4)  = \frac{2 + 9 V_a^2 - 2 \sqrt{1 + 9 V_a^2}}{3 \sqrt{1 + 9 V_a^2}}
\ee
and it is easy to check that, remarkably enough,
\be
K_0(x_c, d =4) > 0
\ee
for $V_a > 0$.  In other words, the instanton has less action that the background and instead of being exponentially suppressed the decay rate is exponentially enhanced.  In general dimension it is also true that
\be
K_0(x_c) > 0
\ee
as can be seen by noting that
\be
\partial_{V_a} K_0 (x_c) = \frac{ (d - 1)^{d-1} \, V_a^{d-1}}{(1+(d-1)^2 \, V_a^2)^{(d-1)/2}} > 0
\ee
and
\be
\lim_{V_a \rightarrow 0} K_0(x_c) = 0
\ee
that is, as the tension wall goes to zero the wall shrinks to zero size and the instanton spends all of its time in the false vacuum.

There does not appear to be a simple expansion in general $d$ for $K_0(x_c)$ in the limit of small tension but for small $V_a$  (\ref{4dflatdec}) becomes
\be
K_0(x_c, d = 4) = \frac{27 V_a^4}{4} + \mathcal{O}(V_a^6)
\ee
In the limit of large tension
\be
x = 1 -\frac{1}{2 (d-1)^2 V_a^2} + \mathcal{O}(V_a^{-4})
\ee
and
\be
K_0(x_c) = V_a - \frac{\sqrt{\pi} \, \Gamma \Big(\frac{d}{2} \Big)}{2 \, \Gamma \Big(\frac{d+1}{2} \Big)} + \frac{1}{2 (d - 1) V_a} + \mathcal{O}(V_a^{-3})
\ee

\subsection{dS to dS decays}

In this case for $0 \leq \tau \leq \tau_0$
\be
\rho' = \cos \omega_0 \tau
\ee
and if $\tau_1 \leq \tau \leq \tau_f$
\be
\rho' = \cos \omega_2 (\tau - \delta) = s_0 \sqrt{1-\frac{\omega_2^2}{\omega_0^2} \sin^2 \omega_0 \tau}
\ee
so the requirement (\ref{deltarho}) that $\rho'(\tau_1) \leq \rho'(\tau_0)$ yields, after a few lines of algebra, the requirement that, provided $\omega_0 < \omega_2$ (i.e. starting, as above, near the true vacuum at $\tau = 0$), $\cos \omega_0 \tau_0 > 0$ but allows either $s_0$.  Intuitively this is just the observation that starting in the true vacuum $\rho$ must be increasing when one approaches the wall, as argued earlier on general grounds.  Defining
\be
x = \cos \omega_0 \tau_0
\ee
and
\be
\gamma = \frac{V_0}{V_2} = \frac{\omega_0^2}{\omega_2^2}
\ee
Then one finds
\begin{eqnarray}
K_0 &=& -\gamma^{1-d/2} x F(\frac{1}{2},1-\frac{d}{2},\frac{3}{2}, x^2) + V_a \gamma^{(1-d)/2} (1-x^2)^{(d-1)/2}  \nonumber \\
&+&s_0 \sqrt{1+\frac{x^2-1}{\gamma}}  F(\frac{1}{2},1-\frac{d}{2},\frac{3}{2}, 1+\frac{x^2-1}{\gamma}) \nonumber \\
&+& \frac{\sqrt{\pi}}{2} \frac{\Gamma\Big(\frac{d}{2} \Big)}{\Gamma \Big(\frac{d+1}{2} \Big) } \Big(\gamma^{1-d/2} -1 \Big) \nonumber \\
\end{eqnarray}
and then the statement that $\partial_x K_0 = 0$ is equivalent to $x = x_c$ where
\be \label{K0deriv1}
\frac{s_0 x_c}{\sqrt{1 + \frac{x_c^2-1}{\gamma}}} = \frac{(d-1) V_a \sqrt{\gamma} \, x_c}{\sqrt{1-x_c^2}} + \gamma
\ee
Making the coordinate change
\be
y_c = \frac{x_c}{\sqrt{1-x_c^2}}
\ee
then (\ref{K0deriv1}) becomes
\be \label{K0deriv2}
\frac{s_0 y_c}{\sqrt{y_c^2+1-\gamma^{-1}}} = (d-1) \sqrt{\gamma} V_a y_c + \gamma
\ee
Squaring both sides of (\ref{K0deriv2}) yields a fourth order polynomial for $y_c$
\begin{eqnarray}
0 &=& (d-1)^2 \gamma V_a^2 y_c^4 + 2 (d-1) \gamma^{3/2} V_a y_c^3 + (\gamma - 1) (\gamma+1+(d-1)^2 V_a^2) y_c^2\nonumber \\
&+&2 (d-1) \sqrt{\gamma} (\gamma - 1) V_a y_c + \gamma (\gamma - 1)
\end{eqnarray}
which can be solved exactly for $y_c$, if not necessarily compactly.   Recalling that $x_c > 0$ and hence $y_c > 0$, then (\ref{K0deriv2}) requires $s_0 = 1$ and plotting the roots one only finds (at least for dimensions from four through eleven) a single real positive root and, evaluated at that root, $K_0(x_c) > 0$.  See Figures 3 and 4.  Hence, just as in the decay to a zero energy vacua, the decay rates for all (positive) values of $V_a$ and $\gamma$ are exponentially enhanced.   

\begin{figure}
\centering
	\includegraphics[scale= 0.80]{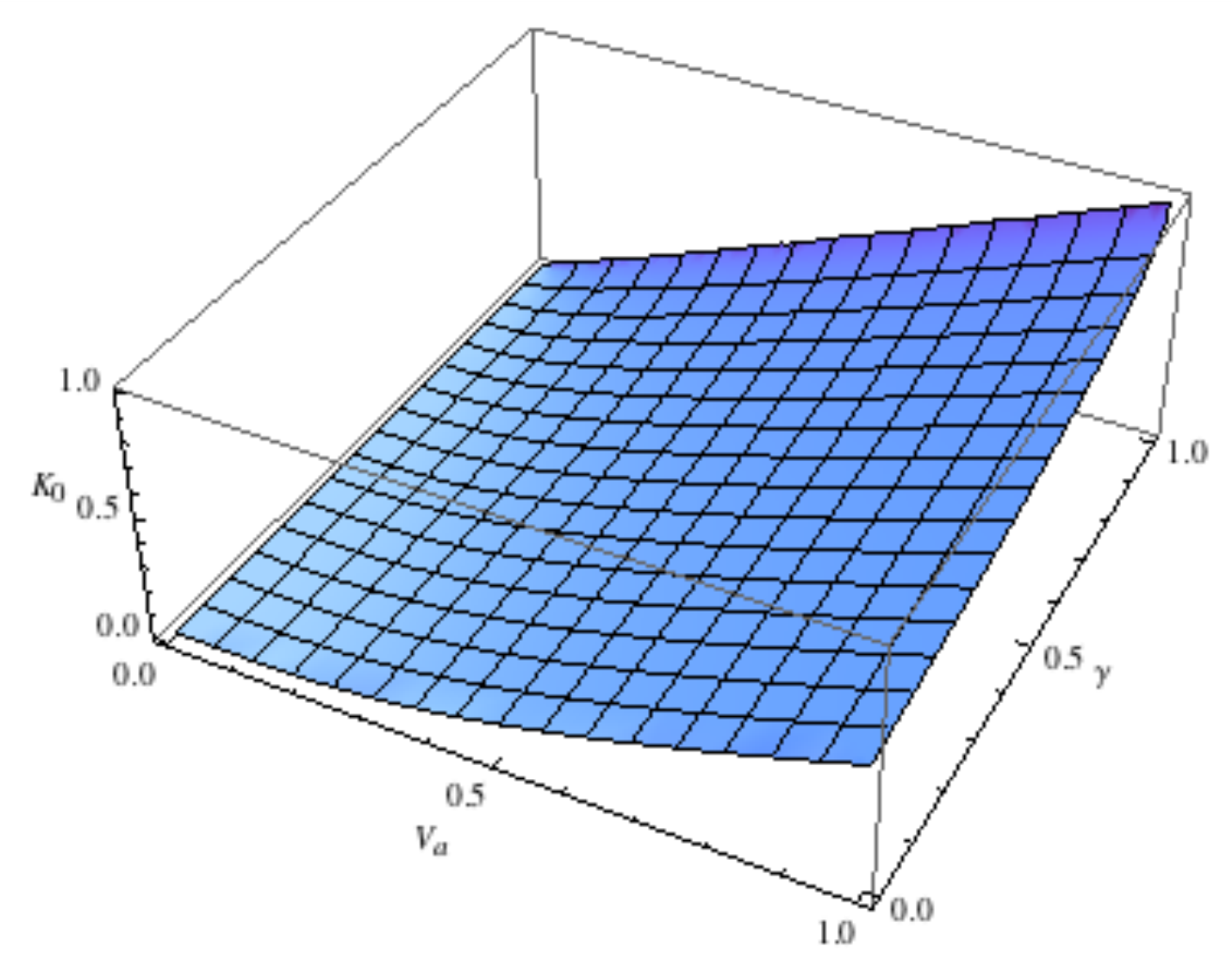}
	\caption{$K_0(x_c, d = 4)$ versus $0 < V_a \leq 1$ and $\gamma$}
	\label{DR1}
	\end{figure}
	
	\begin{figure}
\centering
	\includegraphics[scale= 0.80]{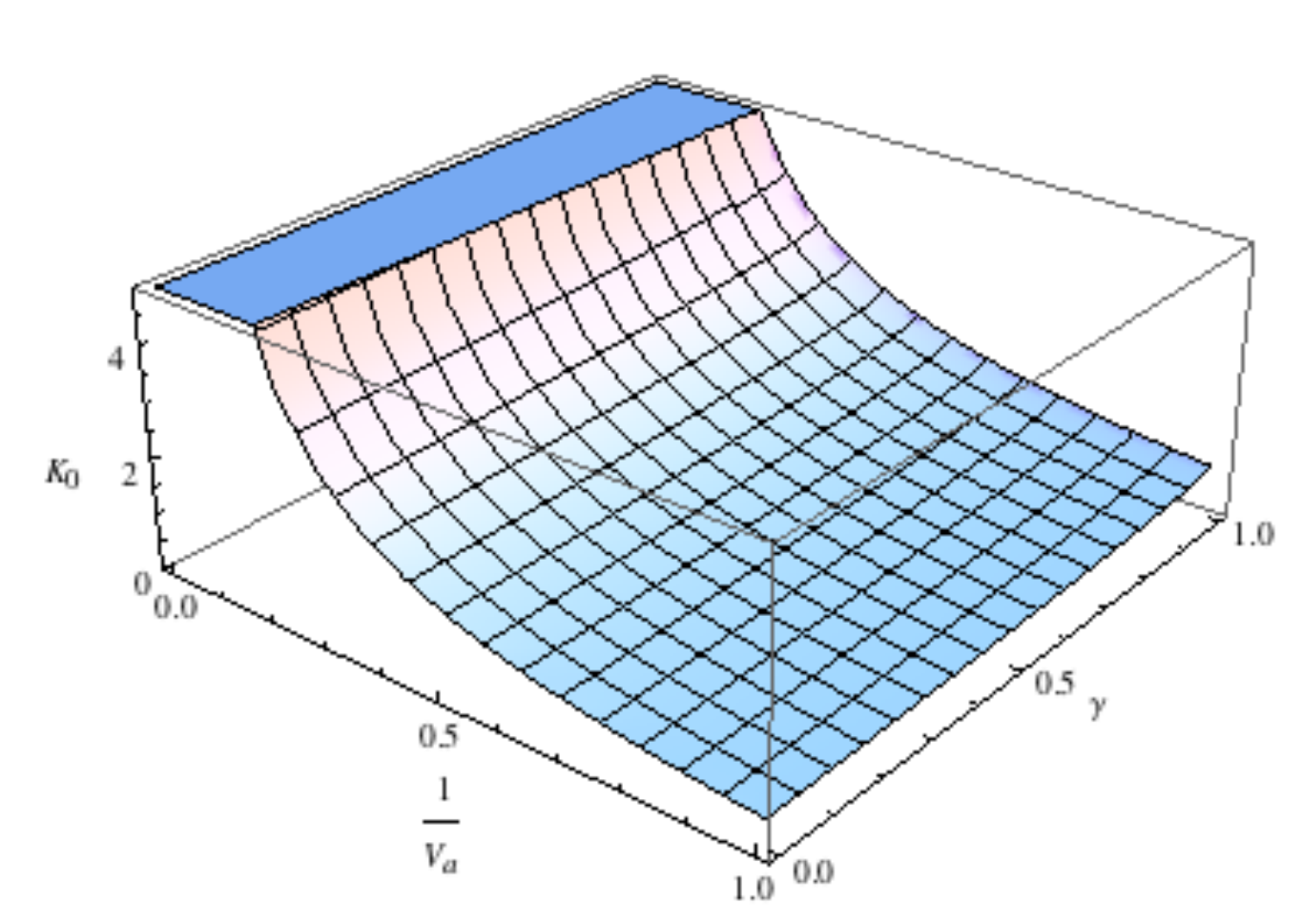}
	\caption{$K_0(x_c, d = 4)$ versus $V_a \geq 1$ and $\gamma$}
	\label{DR2}
	\end{figure}

It is possible, however, to give analytic results for $x_c$ and $K_0(x_c)$ in several limits.  In the limit that the true vacuum potential is small compared to the false vacuum  $\gamma \ll 1$ (i.e. $V_0 \ll V_2$)
\be
x_c = 1 - \frac{(d-1)^2 V_a^2 \gamma}{2 (1 + (d-1)^2 V_a^2)} + \mathcal{O}(\gamma^2)
\ee
and while there does not seem to be a particularly transparent answer for $K_0(x_c)$ in general dimensions for $d = 4$
\be
K_0(x_c, d = 4) = \frac{2 + 9 V_a^2 - 2 \sqrt{1 + 9 V_a^2}}{3 \sqrt{1+9 V_a^2}} + \frac{81 V_a^4 \gamma}{4 (1 + 9 V_a^2)^2}+  \mathcal{O}(\gamma^2)
\ee
and, as expected, the leading order term reproduces the decay rate to flat space.  For nearly degenerate vacua ($\gamma \approx 1$)
\begin{eqnarray}
x_c &=& \frac{(1-\gamma)^{1/3}}{\Big(2 (d-1) V_a\Big)^{1/3}} +\frac{\Big(2 (d - 1) V_a \Big)^{1/3} }{4} (1-\gamma)^{2/3} \nonumber \\
& +& \frac{\Big(2 + (d-1)^2 V_a^2 \Big)}{24 (d-1) V_a}  (1 - \gamma) + \mathcal{O}\Big((1-\gamma)^{4/3}\Big)
\end{eqnarray}
and
\begin{eqnarray}
K_0(x_c) &=& V_a - \frac{3}{4} \Big(2 (d-1) V_a\Big)^{1/3} (1 - \gamma)^{2/3} \nonumber \\
&+&  \Big[ (d-2) \sqrt{\pi} \Gamma \Big(\frac{d}{2} \Big) + (d-1) \Gamma \Big( \frac{d+1}{2} \Big) V_a \Big] \frac{(1 - \gamma)}{4 \Gamma \Big(\frac{d+1}{2} \Big)} + \mathcal{O}\Big((1-\gamma)^{4/3}\Big) \nonumber \\
\end{eqnarray}

In the low tension limit $V_a \ll 1$
\be
x_c = 1 -\frac{(d-1)^2 \gamma}{2 (1-\gamma)^2} V_a^2 + \mathcal{O}(V_a^4)
\ee
and again while $K_0(x_c)$ does not seem to have a simple expression for general $d$, for $d = 4$
\be
K_0(x_c, d = 4) = \frac{27 V_a^4}{4 (1- \gamma)^3} -\frac{243 (1+\gamma)}{4 (1- \gamma)^5} V_a^6+\mathcal{O}(V_a^8)
\ee
and in the high tension limit $V_a \gg 1$
\be
x _c= \sqrt{1 - \gamma} + \frac{\gamma}{2 (d-1)^2 \sqrt{1-\gamma}} \frac{1}{V_a^2} - \frac{\gamma^2}{(d-1)^3 (1-\gamma )} \frac{1}{V_a^3} + \mathcal{O}(V_a^{-4})
\ee
and
\begin{eqnarray}
K_0(x_c) &=& V_a + \frac{\sqrt{\pi}}{2} (\gamma^{1-d/2} - 1) \frac{\Gamma \Big(\frac{d}{2} \Big)}{\Gamma \Big(\frac{d+1}{2} \Big)} - \sqrt{1-\gamma} \gamma^{1-d/2} F(\frac{1}{2}, 1 - \frac{d}{2}, \frac{3}{2}, 1- \gamma) \nonumber \\
&+& \frac{1}{2 (d-1) V_a} + \mathcal{O}(V_a^{-2})
\end{eqnarray}

\subsection{dS to AdS decays}

Turning now to transitions between positive and negative potential vacua, in accordance with the usual convention I will describe these as dS to AdS decays.  It should be noted, however, as stated by Coleman and de Luccia \cite{CdL} and emphasized by a variety of authors since then (e.g. \cite{BanksHeretics, MaldacenaAdSCrunch}), the spacetime after the decay is not global AdS but asymptotically dS with a section which soon runs into a big crunch singularity. In fact this could hardly have been otherwise--AdS is a geodesically incomplete spacetime until a set of boundary conditions is prescribed and it is difficult to understand cosmologically why a given set of boundary conditions (e.g. the usually chosen ``reflecting'' boundary conditions) is chosen by the instanton.

For a true vacuum potential $V(\phi) \approx V_0 < 0$ and a false vacuum potential $V(\phi) \approx V_2$ and defining
\be
\gamma = -\frac{V_0}{V_2}
\ee
then matching $\rho(\tau_0) = \rho(\tau_1)$ leads to the statement that
\be
\sinh \omega_0 \tau_0 = \gamma^{1/2} \sin \omega_2 (\tau_1 - \delta)
\ee
and so 
\be \label{dSAdSmatch}
0 < \sinh \omega_0 \tau_0 < \sqrt{\gamma}
\ee
Note in this case the change in $\rho'$ across the wall
\be
\rho'(\tau_1) - \rho'(\tau_0) = \cos \omega_2 (\tau_1 -\delta) - \cosh \omega_0 \tau_0 
\ee
is always negative and so, unlike the dS to dS case, considering (\ref{Ein2}) does not appear to be terribly useful in ruling out possible solutions.  One might even worry the negative potential region of the wall could make the change in $\rho'$ non-negative, but this possibility is easy to rule out by solving (\ref{Ein1}, \ref{Ein2}) for $\phi'^2$ and integrating $\phi'^2$ from $\tau_0$ to $\tau_1$--the positivity of $\phi'^2$ then shows $\rho'$ must become more negative across the wall.  Using the results in the previous subsections then one finds
\be
I_0 = -\frac{V_2}{\omega_2^d} \gamma^{1-d/2} \int_0^{\omega_0 \tau_0} du \sinh^{d-1} u
\ee
which does not appear to have a simple expression for general $d$ but in even dimensions
\be
I_0 = \frac{V_2}{\omega_2^d} \gamma^{1-d/2} \Bigg[ \cosh \omega_0 \tau_0 F\Big(\frac{1}{2}, 1-\frac{d}{2}, \frac{3}{2}, \cosh^2 \omega_0 \tau_0 \Big) - \frac{\sqrt{\pi} \, \Gamma\Big(\frac{d}{2}\Big)}{2  \, \Gamma \Big( \frac{d+1}{2} \Big)}\Bigg]
\ee
and hence defining
\be
x = \omega_0 \rho(\tau_0) = \sinh \omega_0 \tau_0
\ee
one obtains for even $d$
\begin{eqnarray}
K_0 &=&  (-1)^{d/2} \gamma^{1-d/2} \sqrt{1 + x^2} \, F\Big(\frac{1}{2}, 1-\frac{d}{2}, \frac{3}{2},1+x^2 \Big) +V_a \gamma^{(1-d)/2} x^{d-1} \nonumber \\
 &+& s_0 \sqrt{1-\frac{x^2}{\gamma}} \, F\Big(\frac{1}{2}, 1-\frac{d}{2}, \frac{3}{2}, 1-\frac{x^2}{\gamma} \Big)  - \frac{\sqrt{\pi} \, \Gamma\Big(\frac{d}{2}\Big)}{2 \Gamma \Big( \frac{d+1}{2} \Big)} \Big( (-1)^{d/2} \gamma^{1-d/2} + 1 \Big) \nonumber \\
\end{eqnarray}
and then requiring $\partial_x K_0 = 0$ is equivalent to the statement $x = x_c$ where
\be \label{AdSderivcond}
\frac{s_0 x_c}{\sqrt{\gamma - x_c^2}} = (d-1) V_a -  \frac{x_c \sqrt{\gamma}}{\sqrt{1+x_c^2}}
\ee
or making a change of variables
\be
y_c = \sqrt{\frac{1+\gamma}{\gamma}} \frac{x}{\sqrt{1+x^2}}
\ee
so that (\ref{dSAdSmatch}) is equivalent to $0 < y_c < 1$ and (\ref{AdSderivcond}) becomes
\be \label{AdSderivcond2}
\frac{s_0 y_c}{\sqrt{1- y_c^2}} = (d-1) V_a \sqrt{1+\gamma}  -  \gamma  y_c
\ee
and squaring (\ref{AdSderivcond2})
\begin{eqnarray}  \label{AdSderivcond3}
0 &=& \gamma^2 y_c^4 - 2 \gamma (d-1) V_a \sqrt{1+\gamma} \, y_c^3 + \Big(1+\gamma\Big) \Big(1 - \gamma + (d-1)^2 V_a^2  \Big) y_c^2  \nonumber \\
&+& 2 (d-1) V_a \gamma \sqrt{1+\gamma}  \, y_c - (d-1)^2 V_a^2 (1 + \gamma) \nonumber \\
\end{eqnarray}
Plotting the roots of  (\ref{AdSderivcond3}), for $\gamma \leq 1$(and $V_a > 0$)  one finds a single real root $y_1$ in the relevant range ($0 < y _1 < 1$) and this root is only a solution to the original equation (\ref{AdSderivcond2}) if $s_0 = 1$.   One further finds, much like the dS to dS decays, $K(y_1) > 0 $
and the decay rate is exponentially enhanced rather than exponentially suppressed.  See Figures 5 and 6.  
\begin{figure}
\centering
	\includegraphics[scale= 0.80]{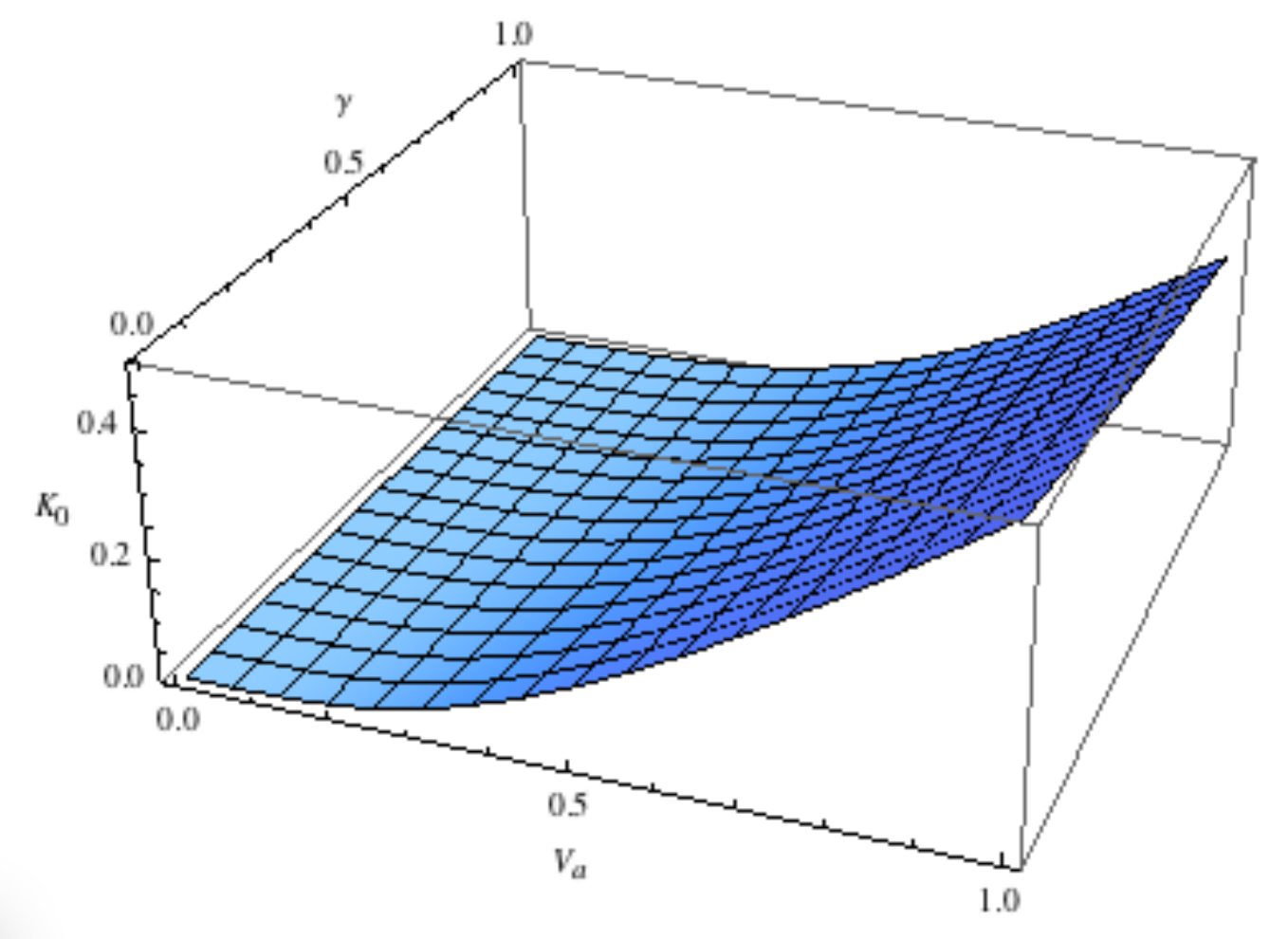}
	\caption{$K_0(y_1, d = 4)$ versus $0 < V_a \leq 1$ and $\gamma$}
	\label{DR3}
	\end{figure}
	
	\begin{figure}
\centering
	\includegraphics[scale= 0.80]{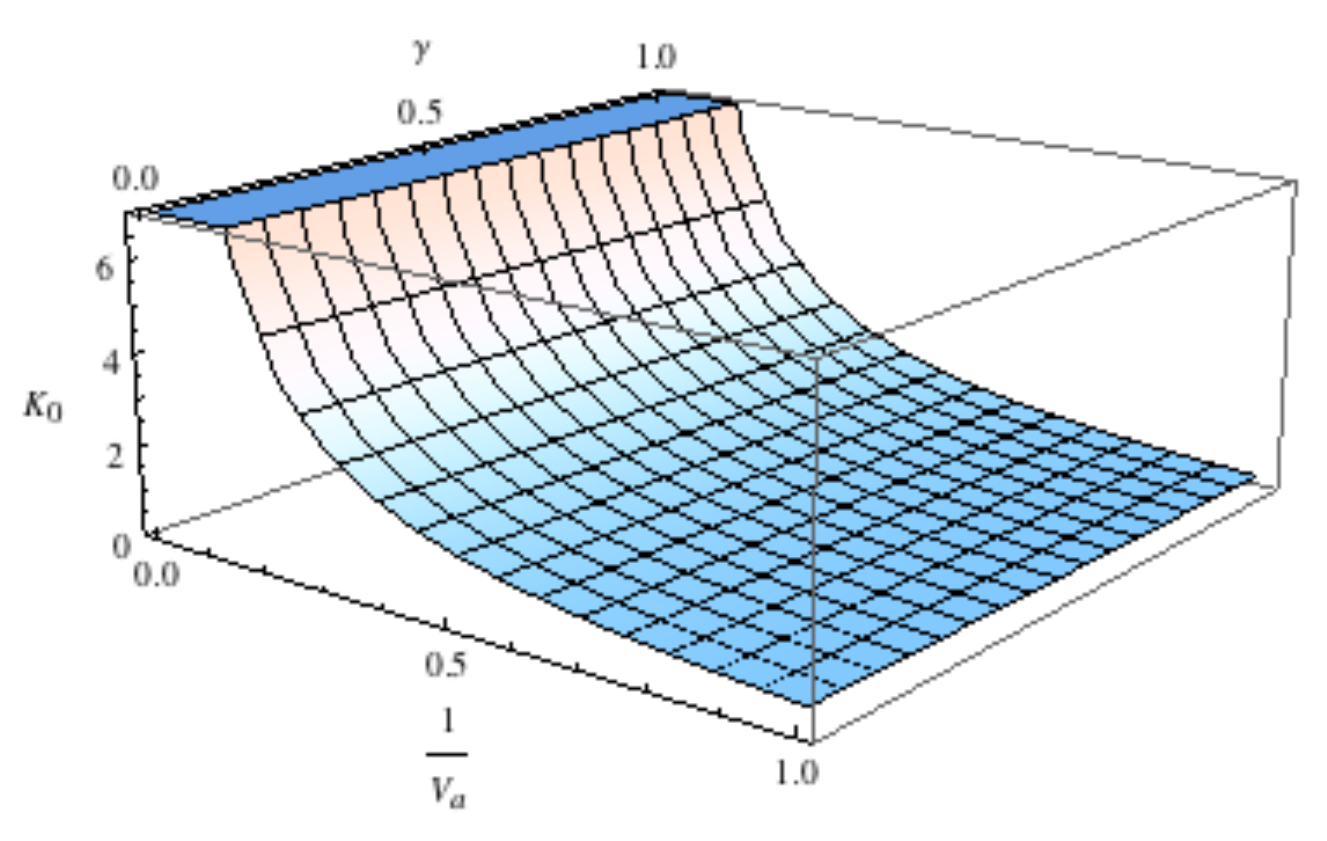}
	\caption{$K_0(y_1, d = 4)$ versus $V_a \geq 1$ and $\gamma$}
	\label{DR4}
	\end{figure}

For $\gamma > 1$ (i.e. the depth of the true vacuum exceeds the height of the false vacuum) and $V_a > 0$, $y_1$ remains real and in the required range and again one finds $K(y_1) > 0$ and very rapid decays.  In addition, however, if $\gamma > 1$ two other roots of (\ref{AdSderivcond3}) become real and are in the relevant ranges in regions of the parameter space $(V_a, \gamma)$.  For these additional roots one has $s_0 = -1$ and $K_0(y_i) < 0$.   There does not seem to be an obvious physical criterion, and in particular those considered above, for rejecting any of these solutions although it is conceivable that not all of these solutions correspond to negative eigenvalues and a true tunnelling interpretation \cite{MarvelTurok}.  In addition, it is possible that, at least for sufficiently large $\gamma$, the thin-wall approximation might break down; the arguments in the previous section assumed the true vacuum was not overly negative.   The existence of multiple solutions would seem to, on the face of it, contradict the argument of \cite{SteinhardtJensenI} regarding the uniqueness of the tunneling instanton and may indicate some of these solutions do not actually occur, for either one of the reasons listed above or due to some complication I have not considered.  I will, however, leave any more detailed analysis of these ``deep well'' instantons for future work.

There does not appear to be an obvious way to rule out negative tension walls in the AdS case and indeed it does not seem unreasonable to expect them if the magnitude of the true vacuum is large compared to the barrier, although I know of no argument showing the validity of the thin-wall approximation in that case.  If one, however, assumes the validity of the thin-wall approximation for negative tension walls and carries out a similiar analysis to the above one finds a single root in the relevant range corresponding to $s_0 = -1$ and $K_0$ evaluated at that root is negative.  That is, while positive tension walls lead to rapid decays, with the possible exception of sufficiently deep true vacuums, negative tension walls lead to slow decays.

\subsection{Flat to AdS decays}

Despite the caveats above regarding various possible subtleties and surface terms for the action  in the asymptotically flat case, as well as the lack of generic justification for the thin-wall approximation for non-thin barriers with these asymptotics, the reader might well regard the failure to discuss flat to AdS decays as a glaring omission.  Hence I will proceed with a treatment of this case under the assumption that none of these difficulties are actually encountered, although these points deserve investigation in future work and conceivably could significantly modify the results in this subsection.  Then I wish to consider bubbles with an interior of negative vacuum energy $V(\phi) = V_0 < 0$ and a false vacuum $V(\phi) = 0$.  In this case, as discussed above, for $0 < \tau < \tau_0$
\be
\rho = \frac{1}{\omega_0} \sinh \omega_0 \tau
\ee
where
\be
\omega_0^2 = -\frac{V_0}{(d-1) (d-2)}
\ee
while for $\tau > \tau_1$
\be
\rho = \tau - \delta_0
\ee
for some constant $\delta_0$ and matching $\rho(\tau_0) = \rho(\tau_1) = \rho_0$ implies
\be
\delta_0 = \tau_1 - \frac{1}{\omega_0} \sinh(\omega_0 \tau)
\ee
Note that for $0 < \tau < \tau_0$, $\rho' = \cosh \omega_0 \tau > 1$ and so the change in $\rho'$ across the wall is necessarily negative and as long as the average of the potential in the wall region is not overly negative there is no (obvious) tension with (\ref{Ein2}).  

In this case $I_2 \approx 0$ since $V(\phi) \rightarrow 0$.  More precisely, as discussed above, $V(\phi)$ will falloff exponentially as a function of $\rho$ asymptotically and $I_2$ becomes arbitrarily small.   The previous writing of the wall tension no longer makes sense in this case so
\be
I_1 = \rho_0^{d-1} \int _{\tau_0}^{\tau_1} d \tau V(\phi) = -\frac{V_0}{\omega_0^d} (\omega_0 \rho_0)^{d-1} V_a
\ee
where
\be
V_a = \omega_0 \int_{\tau_0}^{\tau_1} d \tau \frac{V(\phi)}{(-V_0)}
\ee
As in the last subsection, $I_0$ appears to only have a simple expression for even $d$, so limiting myself to that case and defining
\be
x  = \cosh \omega_0 \tau_0 = \sqrt{1-\omega_0^2 \rho_0^2}
\ee
one finds
\begin{eqnarray}
S_E = -\frac{2 \kappa \Omega_{d-1}}{d-2} \frac{(-V_0)}{\omega_0^d} \Bigg[& (-1)^{d/2}& \Bigg[ x F\Big(\frac{1}{2}, 1 - \frac{d}{2}, \frac{3}{2}, x^2 \Big) - \frac{\sqrt{\pi} \Gamma \Big(\frac{d}{2} \Big)}{2 \Gamma \Big(\frac{d+1}{2} \Big)} \Bigg]  \nonumber \\
&+& V_a (x^2 - 1)^{(d-1)/2} \Bigg]
\end{eqnarray}
and $\partial_x S_E = 0$ at $x = x_c$ implies
\be \label{Flatcond}
0 = -1 + (d-1) V_a \frac{x_c}{(x_c^2-1)^{1/2}}
\ee
which is impossible unless
\be \label{flattens}
0 < V_a < \frac{1}{d-1}
\ee
that is the average of the potential across the wall is necessarily positive but not overly large.   This bound may be understood, as discussed in Coleman-de Luccia \cite{CdL} and rather widely appreciated since, as due to the fact in an AdS region (or more generically any negative energy region) large volumes (compared to the $AdS$ scale) scale as areas and so if the tension of the wall is too positive one can never find the non-trivial zero energy solution needed to produce a decay.   It is worth noting, however, that the lack of a thin-wall instanton does not necessarily imply the solution is in any sense stable; in the case $V(\phi)$ is everywhere negative and only goes to zero asymptotically $V_a < 0$ but that potential in an asymptotically flat space allows solutions with energies which are unbounded from below (or, alternatively, arbitrarily complicated, and hence entropic, solutions of zero energy).

Provided (\ref{flattens}) is met then the action is extremized for $x = x_c$ where
\be
x_c =  (1 - (d-1)^2 V_a^2)^{-1/2}
\ee
and since the on-shell action for the background is identically zero in this case
\begin{eqnarray}
B = S_E(x_c) &=& -\frac{2 \kappa \Omega_{d-1}}{d-2} \frac{(-V_0)}{\omega_0^d} \Bigg[(-1)^{d/2} \Bigg[ x_c F\Big(\frac{1}{2}, 1 - \frac{d}{2}, \frac{3}{2}, x_c^2 \Big) - \frac{\sqrt{\pi} \Gamma \Big(\frac{d}{2} \Big)}{2 \Gamma \Big(\frac{d+1}{2} \Big)} \Bigg]  \nonumber \\
&+& (d-1)^{d-1}  \frac{V_a^d}{(1-(d-1)^2 V_a^2)^{(d-1)/2}}  \Bigg] 
\end{eqnarray}
For $d = 4$ this reduces to
\be
B = -\frac{2 \kappa \Omega_3}{2} \frac{(-V_0)}{\omega_0^4} \frac{2 - 9 V_a^2 -2 \sqrt{1 - 9 V_a^2}}{3 \sqrt{1 - 9 V_a^2}}
\ee
and then it is easy to see $B < 0$ and the decay rate is exponentially enhanced again.  For general even $d$ one can also see $B < 0$ by noting the fact that
\be
\lim_{V_a \rightarrow 0} B \approx 0
\ee
and
\begin{eqnarray}
\partial_{V_a} B &=&   -\frac{2 \kappa \Omega_{d-1}}{d-2} \frac{(-V_0)}{\omega_0^d} (d-1)^{d-1} V_a^{d-1} (1 - (d-1)^2 V_a^2)^{(d-1)/2} \nonumber \\
 &\Big[& (d-1) (1 - (-1)^d) + 1 - (d-1)^2 V_a^2 \Big] 
\end{eqnarray}
and hence $\partial_{V_a} B < 0$.

\setcounter{equation}{0}

\section{Small-wall approximation}

Besides potentials that are accurately described as thin wall there is a second class of potentials that one has some level of calculational control over and expects relatively rapid transitions--those where the difference between the barrier height $V_B$ and the true vacuum $V_T$ is small compared to the values of these potentials.   In other words, in potentials where
\be \label{smallcrit}
\epsilon \equiv \frac{V_B - V_T}{V_T} \ll 1
\ee
and to zeroeth order the potential is constant.   For simplicity of presentation I will assume $V_T > 0$, although an entirely analogous argument goes through for $V < 0$.  If the potential differences the instanton encounters are always small then $\phi'^2$ will never become large and to zeroeth order the action will just be the same as the background.   In other words, simply on grounds of continuity if $\epsilon$ is sufficiently small $B$ will become small and the decay will proceed rapidly.   Then given a densely spaced set of vacua (e.g. the Abbot ``washboard'' potential \cite{Abbotpot}) one should expect the transitions to occur rapidly until (\ref{smallcrit}) breaks down (e.g. typically the last few steps above zero in the Abbot model, depending on the parameters chosen).   More generically, any process involving small energy transitions compared to the effective cosmological constant would fall into this class.

With this background, one may consider potentials
\be
V(\phi) = V_0 + \delta V(\phi)
\ee
where $V_0 \leq V(\phi) \leq V_B$ and $ \delta V(\phi)/V_0 \ll 1$.   One could also just as well use the maximum or average value of the potential to make such a division of the potential,  although this choice will be technically somewhat more useful.   In particular,  $\delta V(\phi) > 0$.  Given the bound on $\phi'^2$ (\ref{philim})
\be
\frac{\phi'^2}{2 V_0} \leq \frac{V(\phi) - V_0}{ V_0 } = \frac{\delta V(\phi)}{V_0} \ll 1
\ee
Then defining $\delta \rho$ so that, as an exact statement,
\be
\rho(\tau) = \frac{1}{\omega_0} \sin \omega_0 \tau + \delta \rho(\tau)
\ee
where, as before,
\be
\omega_0^2 = -\frac{V_0}{(d - 1) (d - 2)}
\ee
then from the Einstein equation
\be
\frac{\rho''}{\rho} = -\frac{-{\phi'}^2}{2 (d - 1)} - \frac{V(\phi)}{(d - 1) (d - 2)}
\ee
one finds
\be \label{deltarho2}
\delta \rho'' + \omega_0^2 \delta \rho \underbrace{\Big(1+ \frac{\delta V}{V_0} + \frac{(d-2)}{2} \frac{{\phi'}^2}{V_0} \Big)}_{= \, 1 + \mathcal{O}(\epsilon)} = -\omega_0 \sin \omega_0 \tau \Big(\frac{\delta V}{V_0} + \frac{(d-2)}{2} \frac{{\phi'}^2}{V_0} \Big) 
\ee
To leading order in $\epsilon$, (\ref{deltarho2}) may be solved with the result that
\begin{eqnarray}
\delta \rho(\tau) &\approx& \cos \omega_0 \tau \Bigg[ K_0 + \int_0^{\tau} ds \sin^2 \omega_0 s  \Big(\frac{\delta V(\phi(s))}{V_0} + \frac{(d-2)}{2} \frac{{\phi'}^2(s)}{V_0} \Big) \Bigg] \nonumber \\
 &+& \sin \omega_0 \tau \Bigg[ K_1 - \int_0^{\tau} ds \sin \omega_0 s \cos \omega_0 s   \Big(\frac{\delta V(\phi(s))}{V_0} + \frac{(d-2)}{2} \frac{{\phi'}^2(s)}{V_0} \Big) \Bigg] \nonumber \\
 \end{eqnarray}
 for some constants $K_0$ and $K_1$.  Recalling that as $\tau \rightarrow 0$ (\ref{phibdy})
 \be
 \rho = \tau + \mathcal{O}(\tau^3)
 \ee
 we are required to set $K_0 = K_1 = 0$ and then find as $\tau \rightarrow 0$, $\delta \rho = \mathcal{O}(\tau^3)$.  Then one finds
 \be \label{drhofin}
 \delta \rho(\tau) \approx \int_0^{\tau} ds \,  \sin \omega_0 s \, \sin \omega_0 (s - \tau) \Bigg[ \frac{\delta V(\phi(s))}{V_0} + \frac{(d-2)}{2} \frac{{\phi'}^2(s)}{V_0} \Bigg]
 \ee
 The instanton ends at $\tau_f$ when $\rho(\tau_f) = 0$ (to be precise, $\tau_f$ is the smallest positive root of $\rho$).   Note that 
 \be
 \rho\Big(\frac{\pi}{\omega_0}\Big) = \delta \rho\Big(\frac{\pi}{\omega_0}\Big) \approx - \int_0^{\pi/{\omega_0}} ds \sin^2 \omega_0 s \Bigg[ \frac{\delta V(\phi(s))}{V_0} + \frac{(d-2)}{2} \frac{{\phi'}^2(s)}{V_0} \Bigg] < 0
 \ee
 since $\delta V(\phi) > 0$.   Hence $\tau_f < \pi/{\omega_0}$.  Note then for $0 < \tau < \tau_f$, $0 < \omega_0 \tau < \pi$ and $-\pi < \omega_0 (s - \tau) < 0$ for $0 < s < \tau$.  Then the integrand of $\delta \rho$ is pointwise negative (\ref{drhofin}) and $\delta \rho < 0$ and further $\delta \rho' < 0$.  Since
 \be
  \delta \rho\Big(\frac{\pi}{\omega_0}\Big) \approx - \frac{1}{\omega_0} \int_0^{\pi} du \sin^2 u \Bigg[ \frac{\delta V(\phi)}{V_0} + \frac{(d-2)}{2} \frac{{\phi'}^2}{V_0} \Bigg]  = \mathcal{O}\Big(\frac{\epsilon}{\omega_0} \Big)
  \ee
  Hence, for $0 < \tau < \tau_f$, $\delta \rho$ is negative and small compared to the scale set by $V_0$, as intuitively expected.  Since
  \be \label{rhofdef}
  \rho(\tau_f) = \frac{1}{\omega_0} \sin \omega_0 \tau_f + \delta \rho(\tau_f) = 0
  \ee
  then
  \be
  1 \gg \sin \omega_0 \tau_f = \sin \omega_0 d\tau
  \ee
  where 
  \be
  \tau_f = \frac{\pi}{\omega_0} - \delta \tau
  \ee
  and either $\omega_0 \delta \tau \ll 1$ or $\omega_0 \tau_f \ll 1$.  The later possibility may be easily eliminated by noting that if $\omega \tau \ll 1$, $\delta \rho$ will be of order $\omega_0^2 \delta \tau^3$ whereas 
  \be
  \frac{1}{\omega_0} \sin \omega_0 \tau = \tau \Big(1 + \mathcal{O}(\omega_0^2 \tau^2) \Big)
  \ee
 for $\omega_0 \tau \ll 1$ and (\ref{rhofdef}) is impossible.   Then it must be true that 
 \be
 \omega_0 \delta \tau \ll 1
 \ee
 Then expanding (\ref{rhofdef}), to leading order
 \be
 \delta \tau \approx \int_0^{\pi} du \sin^2 u  \Bigg[ \frac{\delta V(\phi)}{V_0} + \frac{(d-2)}{2} \frac{{\phi'}^2}{V_0} \Bigg]  
 \ee
 Hence the on-shell action is
 \begin{eqnarray}
 S &=& -\frac{2 \kappa \Omega_{d - 1}}{d - 2} \int_0^{\tau_f} d\tau \rho^{d - 1} V(\phi) \nonumber \\
 &=&  -\frac{2 \kappa \Omega_{d - 1}}{d - 2} \int_0^{\pi/{\omega_0} - \delta \tau} d\tau \Big(\frac{1}{\omega_0} \sin \omega_0 \tau + \delta \rho \Big)^{d - 1} \Big(V_0 +\delta V(\phi) \Big) \nonumber \\
 &=&  -\frac{2 \kappa \Omega_{d - 1}}{d - 2} \frac{V_0}{\omega_0^d} \Bigg[ \int_0^{\pi -\omega_0 \delta \tau} du \sin^{d-1} u + \int_0^{\pi} du \sin^{d-1} u \frac{\delta V(\phi)}{V_0} \nonumber \\
 &+& (d - 1) \int_0^{\pi} du \,  \omega_0 \delta \rho  \sin^{d-2}u + \mathcal{O}(\epsilon^2) \Bigg]
  \end{eqnarray}
  Given the expansion of the integral
  \be
  \int_0^{\pi - x_0} du \sin^{d-1} u = \int_0^{\pi} du \sin^{d-1} u + x_0^{d-1} + \mathcal{O}(x_0^d)
  \ee
  for $x_0 \ll 1$ the contribution of $\delta \tau$ to the on-shell action drops out to first order in $\epsilon$ and rate for tunnelling for these potentials is
\be 
\Gamma = A e^{-B}
\ee
where
\begin{eqnarray}
B  &=& \frac{2 \kappa \Omega_{d - 1}  \Big( (d- 1) (d- 2) \Big)^{d/2}  }{d - 2} V_0^{1-d/2} \int_0^\pi du \sin^{d - 2} u \Big[ \frac{(d-2)}{2} \frac{V_0 - V_F}{V_0} \sin u \nonumber \\
&+& \frac{\delta V}{V_0} \sin u + (d-1) \omega_0 \delta \rho + \mathcal{O} (\epsilon^2) \Big] \nonumber\\
\end{eqnarray}
where $V_F$ is the potential of the false vacuum background and hence
\be
B = \mathcal{O}\Big(\epsilon \, V_0^{1-d/2}\Big)
\ee
Hence if $\epsilon$ is very small in Planck units (e.g. $10^{-120}$) unless $V_0$ is of order $\epsilon^{1/(d/2 - 1)}$ or smaller, $\vert B \vert  \ll 1$ and the decay will be quite rapid.   Even if $\epsilon$ is only moderately small the rate of transition is much faster than one might have naively guessed.   The sign of $B$ is not immediately apparent from this treatment due to the competition of the various terms and may be negative for some, or even all, tunnelling instantons in such potentials.   Note, however, the term due to backreaction ($\delta \rho$) is of the same order of magnitude as the obvious potential contribution and it does not appear clear any non-gravitational treatment of this calculation would yield the same results.

\section{Concluding Remarks}

I have pointed out that rather broad classes of potentials lead to rapid decays of states of perturbatively stable minima once gravity, and some technical subtleties, are taken into account.   If the maxima and minima of the potential are very finely spaced, the above ``small-wall'' approximation applies unless one reaches potentials of order this spacing.  In particular, a densely spaced Abbott washboard-type potential will lead to rapid decays until the last few steps above zero.  It appears likely that such potentials with minima sufficiently close to zero will not decay rapidly via an instanton, this appears to be of little consolation if one is trying to construct minima with energies small compared to typical particle physics scales and in particular of order the present-day cosmological constant--by assumption such potentials also have very small barrier heights and even small thermal fluctuations should overwhelm them.  

I also pointed out that any potentials with thin barriers and nearly degenerate mina, thin and high (compared to the minima) barriers, or barriers which are high and lead to instantons which start very near the minima (in the sense described) lead to instantons well-described by the thin-wall approximation and all such instantons, with the possible exceptions of those to sufficiently deep negative potential minima, lead to decays that are not exponentially suppressed but exponentially enhanced.   Such potentials include, for example, the model KKLT potential \cite{KKLT}.  While it is not entirely clear one should take the precise numerical coefficients for such rapid rates seriously, such decays are certainly not slow and proceed fast enough one should not even describe a configuration in the perturbatively stable minima as a well-defined state.  This difference with the classical Coleman-de Luccia calculation can be traced to a technical subtlety in that analysis where some backreaction effects were implicitly neglected in the calculation of the action and decay rate (specifically by assuming the tunnelling instanton lasts the same amount of Euclidean time as the false vacuum background) and once this oversight is corrected one finds the above very rapid decays.

Note one should not be surprised that this effect has not been noted in the rather substantial literature on numerically constructing tunnelling instantons.   As noted above, to approach the thin-wall limit $\phi$ must start exponentially close to the minima and one requires accordingly exponential accuracy to do a truly controlled numerical calculation of the action.  However,  if one does not take the precautions of verifying in fact one is in the thin-wall approximation or checking the calculation at several different levels of accuracy, it is remarkably easy to perform a numerical calculation that does not look obviously pathological or in contradiction of the Coleman-de Luccia results.

At least naively, the two classes of effective potentials one might expect in a string landscape are those with high (e.g. Planck scale) and thin barriers \cite{Susskind:2003kw} and ones with many relatively small barriers and minima due to breaking of moduli and symmetries at very sub-Planckian scales.   On the other hand, in the light of the above results, it appears to be difficult to construct generic potentials leading to long-lived de Sitter states, let along broad families of such solutions.   The description of decays from zero to negative potential minima is more open to question, as discussed above, but a reasonably broad class of such solutions appear to decay rapidly as well.   As discussed above, there are potentials with high barriers which are of order one width, in Planck units, which are not well-described as thin-wall instantons but it is far from clear such examples are generic.  

One could broadly avoid the above conclusions by constructing potentials which are wide, in Planck units; if the curvature at the top of the barrier height becomes sufficiently small the simple Coleman-de Luccia style instanton  ceases to exist and, it appears, one is generically only left with relatively slow modes of decay (e.g. Hawking-Moss).  However, generically such wide potentials, unless they are finely tuned or have very high barriers, result in extremely light scalar fields.  Such fields are objectionable not only on phenomenological grounds but more importantly, in the absence of special symmetries, should not be expected to be stable quantum mechanically--generally one expects loops effects to generate significant masses for scalars (e.g. of order the supersymmetry breaking scale).   Generating stable and natural very wide and very high barriers sounds like a tall order indeed. 

Then constructing long-lived de Sitter vacua would appear to require either an effective potential which does not have zero or negative potential minima (i.e. is not subject to the usual runaway and decompactification issues) or is not well described as thin-wall or small-wall in any direction in field space, or, for some reason, the decay fails to be well-described by the scalar-gravitational analysis above (e.g. \cite{BanksHeretics}).   Without commenting on some of the relatively controversial positions one might take on the last of these possiblities, two comments are in order in regards to how different forms of matter might change the analysis above.   In any era of non-scalar matter domination (e.g. matter, radiation domination) one can not directly use the above results and it may well be that exponentially enhanced decays, for example, do not occur under such circumstances.   However, simply on grounds of continuity one should expect an analog of the ``small-wall'' approximation above even in this circumstance and a matter field with closely spaced minima to tunnel rapidly, even if the effects on the metric are dominated by other sources.  It would be interesting, however, to investigate these issues in detail. 

 A more difficult problem is to try to incorporate the effects of gravity properly in studying brane nucleation.  While the results of the test-brane calculation are well-known \cite{BoussoPolchinski}, the gravitational effects of a brane are known, to the best of my knowledge, only if there are large number of branes present and one may treat the system classically, in which case one typically has to deal with issues of singularities and horizons.  Note the above results shows, at least with with scalars, naively treating the wall as strictly infinitesimal and only concerning oneself with the junction conditions leads to a serious error in calculating the decay rate.  It would be interesting to be able to perform a similar truly controlled calculation with, for example, branes and fluxes, although it would be quite surprising if one obtained qualitatively different results.

\section*{Acknowledgements}
It is a pleasure to thank N. Turok, N. Afshordi, A. Brown, and M. Johnson for useful discussions.   Research at Perimeter Institute is supported by the Government of Canada through Industry Canada and by the Province of Ontario through the Ministry of Research \& Innovation.

\end{document}